\documentclass[preprint,authoryear]{elsarticle} 
\usepackage[obeyspaces]{url}
\usepackage{graphicx}

\journal{arXiv}

\newcommand{\coo}{\ensuremath{\mathrm{CO_2}~}}

\date{}

\begin{document}

\begin{frontmatter}

\title{The Economics of Variable Renewables and Electricity Storage}

\author[Wegener,DIW]{Javier L\'{o}pez Prol\corref{correspondingauthor}}
\cortext[correspondingauthor]{Corresponding author}
\ead{javier.lopez-prol@uni-graz.at}
\author[DIW]{Wolf-Peter Schill}
\ead{wschill@diw.de}

\address[Wegener]{Wegener Center for Climate and Global Change, University of Graz, Austria}
\address[DIW]{German Institute for Economic Research (DIW Berlin)}

\begin{abstract}
The transformation of the electricity sector is a main element of the transition to a decarbonized economy. Conventional generators powered by fossil fuels have to be replaced by variable renewable energy (VRE) sources in combination with electricity storage and other options for providing temporal flexibility. We discuss the market dynamics of increasing VRE penetration and their integration in the electricity system. We describe the merit-order effect (decline of wholesale electricity prices as VRE penetration increases) and the cannibalization effect (decline of VRE value as their penetration increases). We further review the role of electricity storage and other flexibility options for integrating variable renewables, and how storage can contribute to mitigating the two mentioned effects. We also use a stylized open-source model to provide some graphical intuition on this. While relatively high shares of VRE are achievable with moderate amounts of electricity storage, the role of long-term storage increases as the VRE share approaches 100\%.
\end{abstract}

\begin{keyword}
energy transition, decarbonization, variable renewable energy sources, electricity storage, merit-order effect, cannibalization effect
\end{keyword}

\end{frontmatter}

\section{Introduction}

``Climate change is the mother of all externalities: larger, more complex, and more uncertain than any other environmental problem'' \citep{Tol2009}. Electricity and heat production account for 25\% of global \coo emissions \citep{IPCC2014}, and all energy-related emissions represent almost three quarters of the total \citep{Friedlingstein2019}. Fast and substantial emissions reductions related to energy use are essential for mitigating climate change. Compared to other sectors, the electricity sector offers particularly attractive low-cost mitigation options, given the availability of low-carbon energy technologies \citep{IEA2020ETP}.

The use of renewable energy sources is the major and most promising strategy for fast and deep decarbonization of the electricity sector in terms of sustainability, affordability and availability (abundance of economically viable and secure resources). Beyond today's power sectors, renewable electricity can also be used to substitute fossil fuels in other sectors, such as heating and transport, a strategy often referred to as sector coupling. Due to limited potentials of dispatchable renewable energy sources, such as bioenergy, hydropower, and geothermal, most countries focus on the expansion of variable renewable energy (VRE) sources. These include solar photovoltaics (PV) as well as onshore and offshore wind power. 

Yet integrating high shares of VRE into the electricity sector is not without challenges. Traditional electricity markets were characterized by large, dispatchable electricity generation technologies that adjust production to meet the variable electricity demand at every time. In a deeply decarbonized electricity system with high VRE penetration, however, smaller-scale and distributed VRE generators have to be complemented with options that provide temporal flexibility and geographical balancing. One major option for providing temporal flexibility is electricity storage, which is in the focus of this article. Yet several alternative flexibility options on both the demand and supply sides may complement or partly substitute electricity storage. These include demand-side management, dispatchable generation, and sector coupling. Geographical balancing can also help to address temporal flexibility issues (Section \ref{subsub: competitors}).

\begin{figure}[h]
    \centering
    \includegraphics[width=12cm]{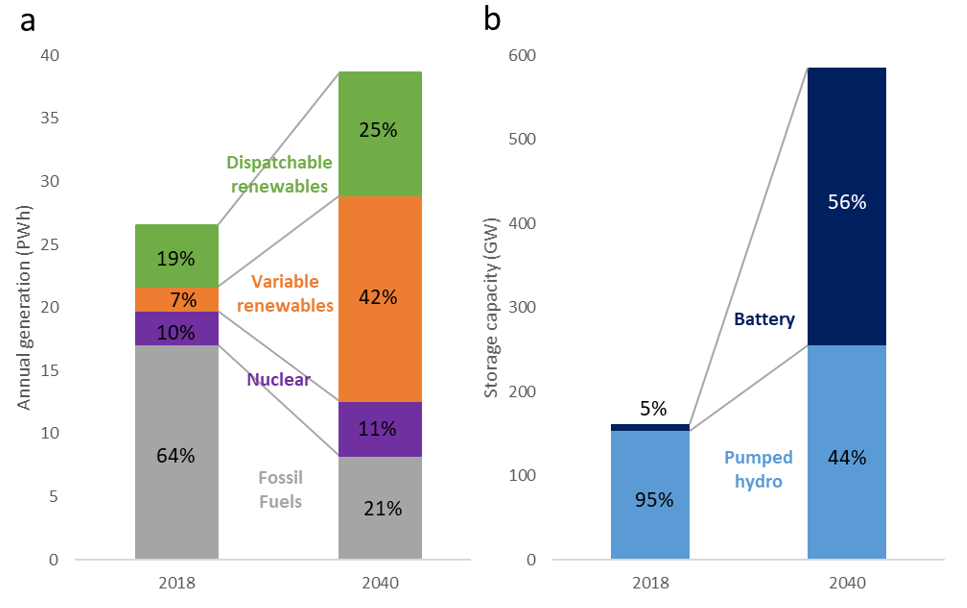}
    \caption{Variable renewables and battery storage are projected to be the primary generation and storage technologies of a decarbonized electricity system. (a) Annual electricity generation (PWh) in 2018 and 2040 of fossil fuels (oil, coal and natural gas), nuclear, variable renewables (wind and solar) and dispatchable renewables (hydro and other renewables) in the Sustainable Development Scenario, \cite{IEA2019WEO}. (b) Installed pumped hydro and battery storage power capacity (GW) in 2018 and 2040 in the Sustainable Development Scenario \citep{IEA2018WEO,IEA2019WEO}.}
    \label{fig: capacities 2040}
\end{figure}

The International Energy Agency (IEA) estimates that in a sustainable development scenario where the Paris Agreement is met, variable renewables should provide 42\% of electricity by 2040, compared to the 7\% of the generation mix they represented in 2018 (Figure~\ref{fig: capacities 2040}). Similarly, whereas the main form of electricity storage is currently pumped hydro, accounting for 95\% of the total installed power capacity in 2018, the IEA estimates that battery storage will grow from a capacity of $8$~GW in 2018 to $330$~GW in 2040, representing thus more than half of the storage power capacity \citep{IEA2018WEO, IEA2019WEO}. Considering that both the IPCC~\citep{Creutzig2017} and the IEA~\citep{Hoekstra2017} tend to underestimate VRE developments in general and solar deployment in particular, these estimations are likely a lower bound of the potential VRE contribution to a decarbonized electricity system. Several studies demonstrate the technical and economic viability of fully renewable electricity supply in Europe \citep{Child_2019} and worldwide \citep{Jacobson_2017}.

Given the prominence of variable renewable energy sources and electricity storage technologies in decarbonized electricity systems, we discuss the economics of VRE, storage, and their interactions. Section \ref{sec: economics VRE} reviews the economics of VRE compared to conventional dispatchable technologies and highlights their electricity market impacts, in particular the merit-order and cannibalization effects. Section \ref{sec: economics of storage and VRE} introduces electricity storage and discusses applications and competing sources of flexibility (\ref{sub: intro storage}), reviews the literature on the interaction between VRE and electricity storage (\ref{sub: literature storage}), and provides graphical intuition on the changing role and value of electricity storage under increasing VRE shares (\ref{sub: intuition storage}).

\section{The economics of variable renewable energy sources}\label{sec: economics VRE}

\subsection{Why variable renewables?}

Fossil fuels (oil, gas and coal) have dominated the energy mix since the industrial revolution, currently accounting for 84.3\% of global primary energy supply and 72.8\% of global electricity generation \citep{BP2020}. Fossil fuels have a high energy density and are suitable for multiple applications, but they have high external costs related to \coo emissions, local pollution, and human toxicity, among others \citep{Ecofys2014, Karkour2020}. 

The Paris Agreement set the goal to reduce \coo emissions to limit global warming ``well below'' $2^{\circ}$C with respect to pre-industrial levels. Given that the earth has already warmed by $1^{\circ}$C \citep{IPCC2018SR1.5}, this goal entails rapid decarbonization of all economic sectors, in which the electricity sector plays a major role. This rapid decarbonization goal entails that current fossil fuel use must be replaced by low-carbon technologies in a relatively short time. To have a perspective of the scale of this challenge, the International Energy Agency estimates that 740~billion~US-\$ will have to be invested annually up to 2030 in clean energy technologies in a sustainable development scenario, compared to 480~billion~US-\$ invested in 2018 \citep{IEA2020Investment}.

\begin{figure}[ht]
    \centering
    \includegraphics[width=12cm]{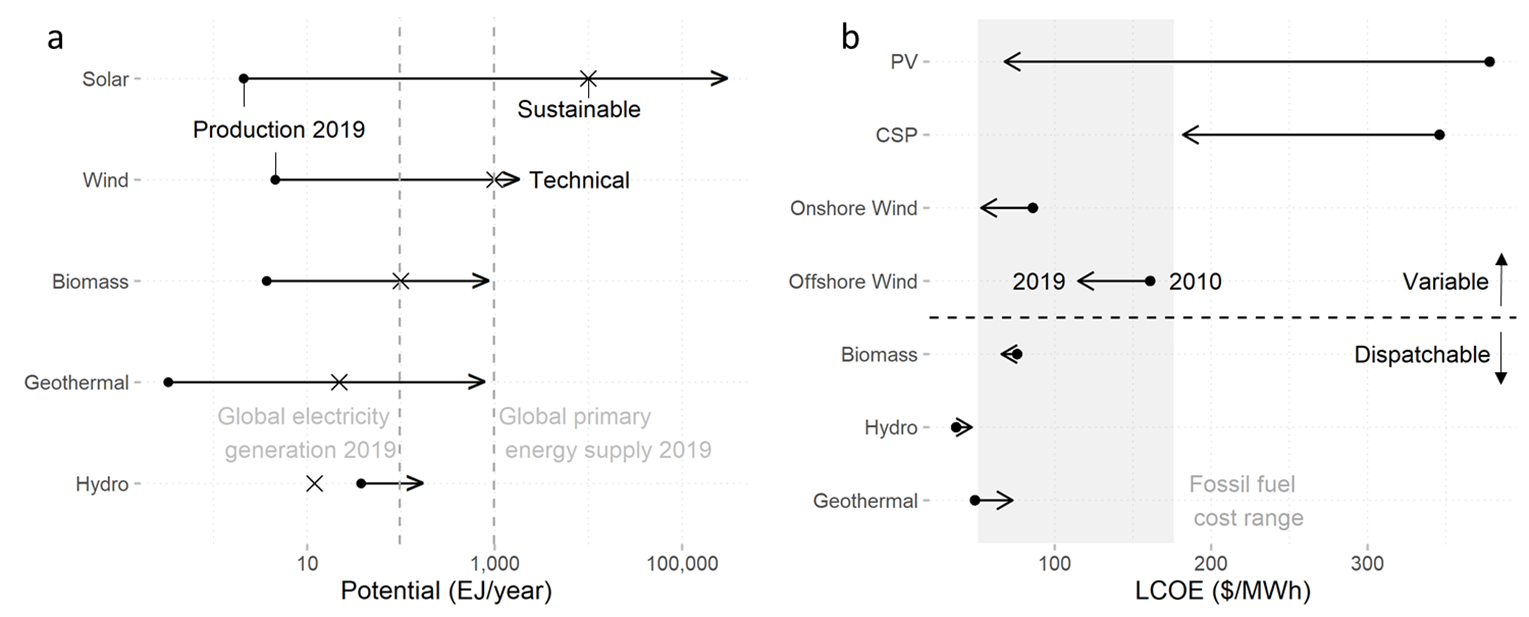}
    \caption{Solar and wind are the most promising renewable energy technologies in terms of availability and cost evolution. (a) Availability (EJ/year): sustainable (cross) and technical (arrow) potential compared to the production of each technology in 2019 (point) and global electricity generation and primary energy supply in 2019 (dashed grey vertical lines). (b) Levelized costs of electricity (US-\$/MWh): average global capacity-weighted unit cost in 2010 (point) and 2019 (arrow) \citep{wbgu2011,IRENA2020}.}
    \label{fig: trilemma}
\end{figure}

Future energy technologies must address three energy policy targets, which are sometimes traded off against each other: sustainability, availability and affordability. Sustainability entails that new energy technologies must be low-carbon and also come with low other external costs, such as biodiversity loss, safety risks, or local pollution. Availability means that energy resources must be abundant, secure and widely accessible. Opposite to fossil fuels or uranium, renewable energy sources provide an annual flow of energy that is available worldwide, and which energy potential exceeds current and prospective global energy needs (Figure~\ref{fig: trilemma}). Finally, affordability requires that the technologies that exploit these resources have a reasonable cost in relation to the value they provide and compared to the alternatives. Wind and solar costs have dropped dramatically in the last decade and are already within the cost range of fossil fuels \citep{IRENA2020}, even without accounting for external costs (Figure~\ref{fig: trilemma}). Additionally, the declining VRE cost is expected to continue \citep{Mayer2015, Agora2017, Vartiainen2020}.

\subsection{Dispatchable vs.~variable technologies}\label{sub: dispatchable vs. variable}

\subsubsection{Dispatchable technologies}
The electricity system is characterized by the constraint that supply and demand must be equal at every point in time. In most markets, electricity demand shows substantial temporal variations, but large parts of the demand do not respond to prices in the short term. Accordingly, electricity supply has to continuously adapt to meet demand. 

Conventional technologies have different degrees of ``dispatchability'', ranging from what is often referred to as ``baseload'' technologies, such as nuclear, that mostly produce evenly across all hours, to ``peak load'' technologies, such as gas turbines, that generally offer higher operational flexibility and only produce during hours of high demand. The dispatchability of a technology is determined by the cost structure of a technology. On top, generation technologies differ with respect to their ability of ramping up and down in the short-term as well as related costs~\citep[cf.~][]{Schill_NE_2017}. Baseload technologies usually have high capital cost and low variable cost, and their output is often relatively inflexible in the short term. Peaking technologies in contrast have lower capital costs and higher variable cost, and they are better able to ramp up or down production in the short term. The composition of the electricity system supply is therefore determined by the operational flexibility and cost structure of the different technologies, which determine their ability to meet demand. This can be summarized in screening curves representing the cost structure of each technology and their equivalent load duration curves showing how each technology covers a segment of demand during a year \citep{Stoft2002}.

\subsubsection{Variable technologies}
The economics of wind \citep{VanKooten2016} and solar \citep{Baker2013} power differ from that of dispatchable technologies \citep[cp.~also~][]{Hirth2016}. The main difference is that variable renewable technologies can only generate when the resource is available, so once a plant is installed, producers do not have any control over production beyond curtailment. They further have a specific cost structure with virtually zero variable cost and relatively high fixed cost. Variability occurs at different timescales and differs for wind power and solar. Very short-term variability, caused for example by clouds for the case of PV, can be easily smoothed out by aggregating installations over larger areas. Diurnal cycles are caused by earth rotation and generally require some form of energy storage. Finally, annual seasonality, caused by earth revolution, is the hardest form of variability as it requires longer-term flexibility across seasons. 

\begin{figure}[h]
    \centering
    \includegraphics[width=12cm]{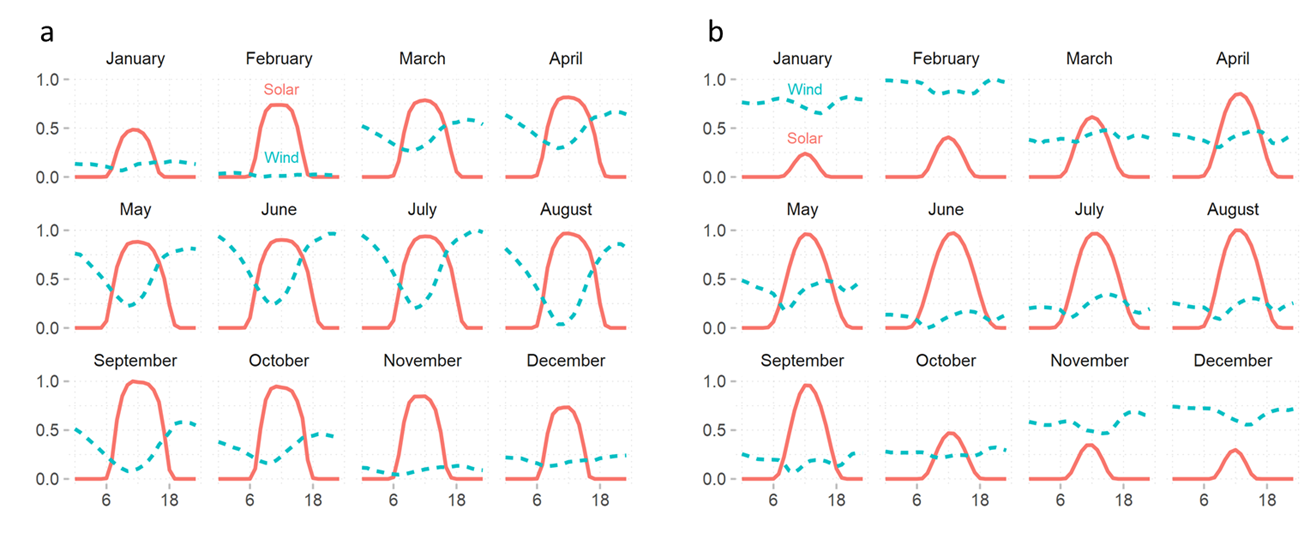}
    \caption{Figure 3. Solar has more homogeneous generation patterns across regions, wind patterns are more location-specific. Normalized average hourly (local times) wind and solar generation per month in California and Germany in 2016. Sources: California data from CAISO adapted from \cite{LopezProl2020}, Germany data from the Open Power System Data platform \citep{Wiese2019}}
    \label{f3} 
\end{figure}

As an exemplary illustration of variable renewable patterns, Figure~\ref{f3} shows normalized (0-1 for each region and technology) wind and solar PV generation in California (Figure~\ref{f3}~a) and Germany (Figure~\ref{f3}~b) during the year 2016. Short-term variability is smoothed out by the country/state-wide spatial aggregation and averaging of the hours of each month (in local time). Solar cycles are homogeneous across regions: solar always generates only by day and more in summer than in winter. Since solar seasonality is smoother as we approach the equator, differences between summer and winter generation are smaller in California than in Germany. Wind power is more location-specific: whereas wind produces more during the night than at noon in both regions, the seasonal peak is in summer in California and winter in Germany.

\subsubsection{The value of dispatchable vs.~variable generation}
Variable renewables do not only differ from dispatchable technologies in terms of costs, but also in terms of value \citep{Joskow2011a}. \cite{Lamont2008} provides an analytical formulation of the difference between the marginal values of VRE and dispatchable technologies. The marginal value ($MV_{D}$) of a dispatchable generator is the sum of the difference between the hourly electricity price ($\lambda_h$) and the generator’s variable cost ($VC$) (assumed constant) for all the hours when the generator produces at its maximum capacity ($h*$).

\begin{equation}
MV_{D} =\sum_{h*} (\lambda_h-VC)
\end{equation}

The marginal value of a variable generator is the average electricity price ($\overline{\lambda}$) times the average capacity factor (actual production as a fraction of peak capacity, $\overline{CF}$) plus the covariance between the hourly price and hourly capacity factor ($Cov(\lambda,CF)$) for all hours of the year ($H$). The first summand represents the average unit revenues of the generator, whereas the second captures its load-matching capability, i.e.,~the correlation between demand and supply patterns.

\begin{equation}
MV_{VRE} = [\overline{\lambda} \cdot \overline{CF} + Cov(\lambda,CF)] \cdot H 
\end{equation}

Whereas the marginal value of dispatchable technologies depends on a random variable (price) and constant (variable cost), the marginal value of VRE depends on the covariance between price and capacity factor. Since VRE have zero variable cost, increasing VRE penetration reduces electricity prices due to the merit-order effect (MOE). The covariance between price and capacity factor depends on the share of VRE, as the price will generally be lower the higher VRE penetration, all other factors held constant. The merit-order effect (reduction of prices as VRE penetration increases, see section~\ref{sub: merit-order effect}) and the cannibalization effect (reduction of VRE value as penetration increases, see section~\ref{sub: cannibalization effect}) have been theoretically explained in the literature and empirically estimated in countries with relatively high VRE penetration. The next sections introduce a graphical conceptualization of these phenomena and review the corresponding empirical findings.

\subsection{The merit-order effect}\label{sub: merit-order effect}

Electricity is a perfectly homogeneous good in three dimensions: time, space, and lead-time between contract and delivery \citep{Hirth2016}. This entails that the market clears at marginal cost for every time, location and market run, but the price may substantially vary across these three dimensions, depending on the market design. Since VRE have zero variable cost, higher penetration of VRE depresses wholesale electricity prices. The equilibrium price and quantity of electricity delivered each hour to the market is given by the intersection between demand and supply. 

\begin{figure}[h]
    \centering
    \includegraphics[width=12cm]{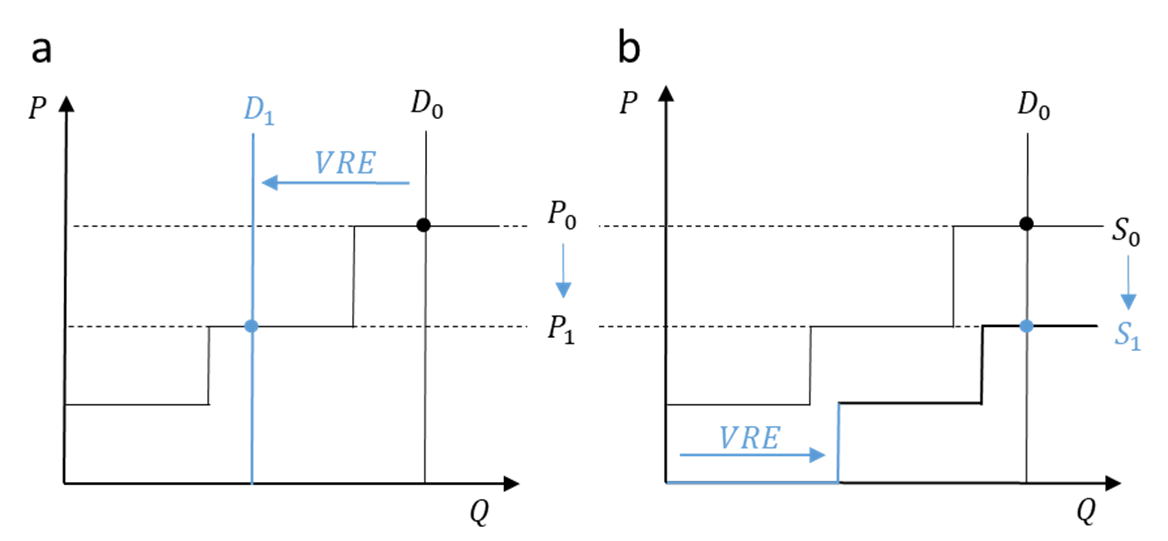}
    \caption{Figure 4. The merit-order effect of variable renewables in wholesale electricity markets. (a) Demand perspective. (b) Supply perspective.}
    \label{f4} 
\end{figure}

Demand is usually assumed to be perfectly inelastic in the short term, as most consumers so far do not respond to real-time prices. For this reason, it is represented as a vertical straight line in Figure~\ref{f4}. This is not the case in forward markets, as consumers can plan ahead their consumption, and bid accordingly. Additionally, demand elasticity is likely to increase in the future in most markets, as different measures of demand response are likely to be implemented to provide flexibility to the electricity system. Still, we represent demand as perfectly inelastic in Figure \ref{f4} for simplicity. 

The supply curve is given by the variable cost (or more precisely, the opportunity cost of fuel) of all power plants sorted in ascending order from lower (left) to higher (right). Figure~\ref{f4} represents a simplified supply curve with three steps representing three technologies with different variable costs. The technology with lower variable costs on the left-hand side of the supply curve could be referred to as a baseload technology, such as lignite or nuclear. The middle step represents a mid-load technology, such as hard coal or combined-cycle gas turbines, and the last step of the supply curve represents a peaking technology, such as open-cycle gas or oil turbines.

In this basic setting of a wholesale electricity market, VRE generation can be interpreted as a lower residual demand, also referred to as net load, (Figure~\ref{f4}~a) or as higher zero variable cost supply (Figure~\ref{f4}~b). From a demand perspective, VRE generation is represented as lower net load. The demand curve shifts to the left ($D_0 \rightarrow D_1$), and therefore the price declines ($P_0 \rightarrow P_1$) as the technology setting the price is now the lower-cost mid-load rather than the peak-load technology (Figure~\ref{f4}~a). From a supply perspective, VRE generation shifts the supply curve to the right as new zero variable cost is supplied to the market ($S_0 \rightarrow S_1$). The same price effect comes into play, lowering wholesale electricity prices from $P_0$ to $P_1$.

The dynamics of the merit-order effect are well explained in the electricity market economics literature \citep{Stoft2002, Green2015, Leautier2019}, and the effect also has been widely studied in the numerical and empirical literature. The numerical literature provides ex-ante estimations of the MOE with different types of electricity market models. The advantage of such market simulations is that they provide insights on potential future scenarios before they actually happen. Some early examples of this type of ex-ante modelling exercise to quantify the merit-order effect are \cite{SaenzdeMiera2008}, who estimate the MOE using a dispatch model in Spain, \cite{Sensfuss2008}, who apply an agent-based model to Germany, and \cite{Mills2012a}, who use a dispatch and investment model for California.

As VRE penetration increases, empirical ex-post studies with actual historical market data estimate the MOE in wholesale electricity markets. Most studies find evidence of the depressing effect of VRE penetration on wholesale prices in countries with high penetration such as Germany \citep{Cludius2014}, Italy \citep{Clo2015a}, Spain \citep{Gelabert2011}, Australia \citep{Csereklyei2019}, California \citep{Woo2016}, and Texas \citep{Woo2011}. A cross-country analysis of the main European power markets shows consistent negative results across countries \citep{Welisch2016}. Likewise, a review of the MOE in the USA confirms the depressing effect of VRE on wholesale electricity markets, although this effect is relatively small compared to the effect of declining gas prices \citep{Mills2020}, which decreased significantly since 2008. In general, the MOE of solar is stronger than that of wind as its generation pattern is more concentrated during day-time hours (see Figure~\ref{f3}).

\subsection{Price volatility}

Given the inherent variability of VRE, it is generally understood that increasing VRE penetration increases wholesale price volatility. \cite{Seel2018} find that wholesale price variability increases as solar and (to a lesser extent) wind generation increases across the main US electricity markets using a capacity expansion and unit commitment model.

The empirical evidence on the effect of VRE penetration on price volatility is however mixed. The effects are specific to (i) the considered timeframe (hourly, daily or weekly volatility), (ii) the VRE technology (solar or wind have different generation patterns and therefore differing effects), and (iii) the conditions of the electricity system itself (market design, availability of flexibility options, demand patterns, etc.). For instance, \cite{Rintamaki2017} find that whereas weekly volatility increases in both Germany and Denmark due to increasing wind and solar penetration, the daily volatility patterns differ. Wind decreases daily volatility in Denmark, but increases it in Germany, whereas solar decreases daily volatility in Germany. Similarly, \cite{Kyritsis2017} find that solar energy decreases price volatility in Germany by scaling down the use of peak-load power plants, whereas wind increases the volatility due to increased flexibility needs. Both \cite{Rai2020} in Australia and \cite{Ballester2015} in Spain find that VRE increases volatility, but reduces the persistence of price spikes. Finally, \cite{Ciarreta2020} show that market design and VRE regulation choices may reduce price volatility even as VRE penetration increases. 

\subsection{The cannibalization effect}\label{sub: cannibalization effect}

Whereas an increase in supply causes a decline in prices in any market, the cannibalization effect is a specific phenomenon of VRE penetration in current electricity markets. This is due to specific properties of the good electricity and the characteristics of VRE. 

Since electricity is a perfectly homogeneous good with relatively inelastic, time-varying demand and relatively high bulk storage costs, it is governed by peak-load pricing. Peak-load pricing theory entails that off-peak prices reflect short-run marginal operational costs, whereas fixed costs are recovered through peak-load scarcity prices \citep{Boiteux1949, Wenders1976}. Whereas energy-only markets in theory reflect peak-load pricing, most real-world electricity markets cap prices, such that fixed costs of peak generators are usually recovered through some kind of explicit or implicit capacity payment rather than through scarcity prices \citep[cf.][]{Cramton_2012, Cramton2017}.

Variable renewables have virtually zero short-run marginal cost and their generation potential varies over time. Together with the constraint that supply and demand have to be balanced at every time, these conditions entail that prices drop whenever VRE generation enters the market. Depending on the existence and design of renewable support schemes, as well as on flexibility constraints of other generators, prices can even become negative when VRE supply exceeds demand.

\begin{figure}[h]
    \centering
    \includegraphics[width=12cm]{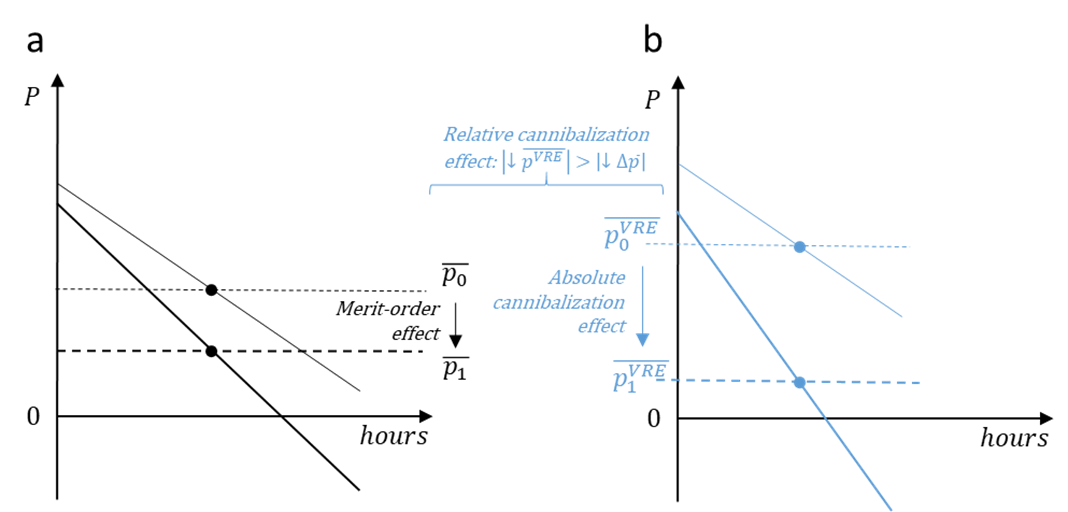}
    \caption{Figure 5. The merit-order and the cannibalization effect. Price duration curves for the electricity market (a) and for a variable renewable generator (b) with low ($_0$) and high ($_1$) VRE penetration.}
    \label{f5} 
\end{figure}

The cannibalization effect can be illustrated through stylized price duration curves, representing prices in the wholesale market for all the 8760 hours of a year in descending order. Figure~\ref{f5}~a represents the merit-order effect observed in the wholesale electricity market. As VRE penetration increases, prices decline (the price duration curve shifts downwards) and the average wholesale electricity price declines from $\overline{p_0}$ to $\overline{p_1}$. 

Figure~\ref{f5}~b represents a stylized price duration curve for a VRE technology such as solar PV, i.e,~the price received by this VRE generator during every hour of production. Solar is typically correlated with electricity demand and therefore benefits from higher peak-time electricity prices. For this reason, the unit revenue (generation-weighted electricity prices, also called market value) received by solar generators at low penetration is higher than the average wholesale electricity price in this example.

However, as solar penetration increases, the merit-order effect depresses electricity prices. Since the merit-order effect happens primarily during the hours of high solar energy availability, the unit revenues received by solar PV fall $\overline{p^{VRE}_0} \rightarrow \overline{p^{VRE}_1}$, stronger than the average wholesale electricity price. The absolute cannibalization effect is the fall of VRE unit revenues. The relative cannibalization effect refers to the fact that VRE unit revenues fall faster than average wholesale electricity prices, and is represented by the fraction between both. This can also be expressed using the value factor concept. A value factor of 1 (or 100\%) means that the unit revenues received by VRE generators are exactly the same as the average wholesale price. The value factor is smaller than 1 when VRE unit revenues fall more than average wholesale prices. Additionally, higher VRE penetration makes the price duration curves steeper \citep{Green2011} because VRE increases the difference between hours with low and high prices. When VRE are producing, the merit-order effect depresses prices. However, given the increased flexibility needs of the electricity system, prices may increase in other times due to the required ramp-up of more flexible generators with higher marginal cost \citep{Bushnell2018}.

As with the merit-order effect, there is a rich literature estimating the declining value of variable renewables both ex-ante with numerical models and ex-post with econometric methods. Ex-ante dispatch and investment models are used to estimate the optimal penetration of variable renewables under different scenarios \citep{Hirth2015d} and the declining market value of wind and solar in different markets and scenarios \citep{Hirth2013a, Mills2020, Eising2020}. Additionally, Mills and Wiser (\citeyear{Mills2014}) estimate the potential of different strategies to mitigate the cannibalization effect of wind and solar. They find that geographic diversification is the most promising strategy to mitigate the declining value trend of wind power at high penetration, and low-cost storage for PV. The mitigating effect of storage and flexibility is discussed in more detail in Section \ref{sec: economics of storage and VRE}. 

From an empirical perspective, several studies report the declining value of wind and solar technologies \citep{Zipp2017, Welisch2016}. \cite{Clo2015} show that increasing solar penetration reduces its own value but increases the value of gas in the Italian wholesale electricity market. \cite{LopezProl2020} define and estimate the absolute and relative cannibalization and cross-cannibalization (between VRE technologies) effects in the California wholesale electricity market. Whereas wind decreases its own and solar value, both in relative and absolute terms, solar decreases its own market value and value factor, but increases the wind value factor. Given the findings of \cite{Bushnell2018}, this might be because when the sun sets, more flexible generators with higher marginal cost are necessary to cover declining solar production, thus increasing wholesale prices at the time when wind generation increases.

The literature agrees that increasing VRE penetration decreases their market value (absolute cannibalization) at a faster pace than the fall of wholesale prices (relative cannibalization) due to the merit-order effect. Whereas ex-post estimations show that the cannibalization effect increases for higher VRE penetration \citep{LopezProl2020}, ex-ante market models shows that market value tends to stabilize as penetration increases \citep{Hirth2013a}. This reflects the different time frames of both types of methods: whereas econometric models estimate the cannibalization effect in the short term in the existing electricity system, numerical capacity expansion models optimally adapt installed capacities to cope with higher shares of VRE. 

\subsection{A value perspective of integration costs}

Integrating VRE into current electricity systems that have been designed for dispatchable generation technologies entail integration costs derived from the variability (profile costs), uncertainty (balancing costs) and location specificity (grid costs) of VRE resources \citep{Ueckerdt2013a}. \cite{Hirth2015e} show that integration costs can be estimated from a cost perspective as the additional system costs caused by VRE integration, or from a value perspective as the difference between wholesale prices and VRE market values. Since the value factor is the ratio between a technology's market value (unit revenues) and wholesale electricity prices, the evolution of the value factor (the relative cannibalization effect) can be interpreted as a value estimation of the technology-specific integration cost.

\section{The economics of electricity storage for VRE integration}\label{sec: economics of storage and VRE}

\subsection{Why electricity storage?}\label{sub: intro storage}

\subsubsection{General effects of electricity storage}

The use of electricity storage is an important option for addressing the variability challenges of wind and solar power discussed above. Electricity storage may be broadly defined as a technology that takes up electrical energy at a given point in time, and releases electrical energy again at a later point in time, irrespective of the form in which energy is stored in between. Electricity storage can balance the variable patterns of renewable supply and demand and thus mitigate both the merit-order and the cannibalization effects.

\subsubsection{Technologies}

A wide range of electricity storage technologies is available \citep{Luo_2015, WEC2016, Guer_2018}. These differ with respect to various techno-economic parameters, such as round-trip efficiency, specific investment costs for power and energy capacity, variable costs, and technical lifetime. Driven by these parameter differences, typical energy-to-power (E/P) ratios emerge for each technology, the inverse of which is also referred to as ``C-rate''. For example, the E/P ratio is only a few hours for stationary Li-ion batteries, may exceed ten hours for pumped hydro storage plants, and can surpass several weeks for hydrogen-based electricity storage. Further, electricity storage technologies come at very different scales. For example, stationary batteries are highly modular and can be deployed at the scale of individual households with energy capacities of only a few kWh, but also at the grid scale in the multi-MWh range. Pumped hydro plants, in contrast, are usually both lumpier and much larger, with energy capacity in the GWh range. Future hydrogen storage may well reach the TWh range.

In recent years, substantial research and development activities as well as scaling up of production has led to substantial cost decreases of electrochemical storage technologies, which are expected to continue also in the future \citep{Schmidt_2017, Lazard2019a, Schmidt2019}. In the medium run, Li-ion batteries may become the dominating storage technology \citep{Beuse2020}. Various storage technologies are already available and technologically proven today, although they are not necessarily economic in current power markets \citep{Brown_response_2018}. This is partly because of economic and regulatory barriers in many markets, which prevent storage from being fully compensated for the many services it could deliver to the electricity system \citep{Sioshansi_2012, Castillo_2014}.

\subsubsection{Applications}
There are many different applications for electricity storage \citep{Palizban_2016}. A well-known and widely used application is bulk electricity storage, i.e.,~shifting substantial quantities of electricity from one point in time to another point in time. This business model is also referred to as energy arbitrage in wholesale electricity markets, which entails buying and storing electricity in periods of low prices and selling it again in higher-price periods. In doing so, the price difference has to pay for the variable storage costs, including the roundtrip losses. From an overall system perspective, bulk electricity storage allows increasing the use of generators with low variable costs and decreasing the use of generators with high variable costs. With increasing shares of VRE, this becomes more relevant, in particular in case of temporary renewable surplus generation which would otherwise have to be curtailed \citep{Denholm_2013}.

Aside from energy arbitrage, there are many other potential applications for electricity storage, which are not necessarily organized by markets \citep{Denholm_2010, Castillo_2014}. These include: (i) reduction of residual load gradients, which translates into reduced ramping needs of thermal generators; (ii) provision of different types of ancillary services, in particular balancing power; (iii) provision of peak or firm generation capacity; (iv) supplementation or deferral of investments into transmission or distribution grid infrastructure; (v) multiple end-user applications, including energy supply cost optimization and power quality or stability. These applications can be translated into different types of value created by electricity storage in the energy sector, such as arbitrage, flexibility, reserve, capacity or grid-related values. Importantly, storage facilities may be able to combine several of these applications at least to some extent \citep{Stephan_2016, Staffell_2016}. Many model-based studies nonetheless focus on only one or two sources of storage value, and hardly any cover all of them.

\subsubsection{Alternative flexibility options}\label{subsub: competitors}

Aside from electricity storage, there are many other, and partially competing, options for providing temporal flexibility to the electricity sector \citep{Lund_2015}. These belong either to the supply side or to the demand side and may be grouped into different categories. Depending on electricity flowing into and/or out of the flexibility option, we distinguish the categories Power-to-Power, X-to-Power, or Power-to-X, following \cite{Schill_2015} and \cite{Schill_Joule_2020}. Electricity storage by definition belongs to the Power-to-Power category.

\textbf{Power-to-Power:} Demand-side management measures can have similar electricity sector impacts as electricity storage \citep{Hale2018}. If a part of the electricity demand is shifted from hour~$n$ to hour~$n+1$ (by decreasing the load in hour~$n$ and correspondingly increasing it in hour~$n+1$), this has a similar power sector effect as discharging an electricity storage facility in hour~$n$, and charging it again in hour~$n+1$. Realizing demand-side flexibility may require digitalization \citep{IEA2020Digital} to automatize demand response to prices, and market design changes to allow agents to capture the value they generate to the system by providing flexibility \citep{Joskow2019, Leslie2020}. For clarity, here we only include shifting of such electricity demand that already exists in traditional power sectors, e.g.,~in residential applications or industrial processes, and not additional future loads related to increased sector coupling.

\textbf{X-to-Power}: This category includes any technologies that generate electricity at the time of demand. This comprises conventional dispatchable generators, and also measures to operate flexibility-constrained assets such as combined heat and power generation plants in a more flexible way. VRE overcapacity with curtailment may also be categorized here.

\textbf{Power-to-X}: This group of options includes new and potentially flexible electric loads related to the coupling of the power sector and other sectors. Such sector coupling is a major option to facilitate deeply decarbonized future energy systems \citep{IPCC_2018}. For example, electricity can be used in the heating sector via heat pumps or direct resistive heating \citep{Bloess_2018, Madeddu_2020}, in the mobility sector via battery-electric vehicles \citep{Armaroli_2011}, or in various transportation or industrial applications, which are harder to electrify, via electrolysis-based hydrogen or synthetic liquid fuels \citep{Schiebahn_2015, Mansilla2018, Runge_2019, IEA2019H2}. Such sector coupling allows making use of other, and potentially cheaper, energy storage options than stationary electricity storage, such as thermal energy storage, electric vehicle batteries, or storage of hydrogen or derived synthetic liquid fuels. Note that we define Power-to-X as any process where electricity permanently flows out of the power sector. In case of later reconversion to electricity, e.g.,~if electricity is generated from hydrogen or synthetic liquid fuels again, this is treated as a Power-to-Power option \citep[cp.][]{Yan_2019}. Battery-electric vehicles that discharge electricity back to the grid (``Vehicle-to-Grid'') provide both Power-to-X (the part of electricity used for driving) and Power-to-Power (the part of electricity fed back to the grid) functionalities \citep[cp.][]{Taljegard_2019}.

Geographical balancing is another option to provide temporal flexibility for variable renewable integration\citep{Fuersch_2013, Steinke_2013}, as illustrated for the US \citep{MacDonald_2016} and Europe \citep{Schlachtberger_2017}. This is because the time patterns of renewable availability and electricity demand are usually not perfectly correlated between different locations \citep{Engeland_2017}. Transmission expansion can thus facilitate balancing of renewable supply and demand over wide areas. In theory, annual and diurnal solar cycles could be balanced using inter-hemispheric connections to offset seasonal patterns \citep{Grossmann2014, Aghahosseini2019}, yet the practical feasibility of this option remains unclear.

\subsection{Modeling the interaction between storage and variable renewables}\label{sub: literature storage}

A broad literature has investigated the role and value of electricity storage in electricity sectors with variable renewable energy sources. Analyses vary not only in terms of the geographical coverage and time horizon, but also with respect to the methodological approaches used \citep{Zerrahn_2017, Blanco_2018}. Several earlier studies use price-taking arbitrage models, drawing on historic electricity market prices. Another strand of the literature uses models based on time series of variable renewable availability and load; and a third strand uses electricity sector models, often with a capacity expansion approach. The latter may be considered the state-of-the-art approach to quantify the value of electricity storage in future scenarios with very high shares of variable renewables. \cite{Bistline_2020} review and discuss the representation of energy storage in respective long-term models.

\subsubsection{Price-taker arbitrage models}
Several studies have investigated the value of electricity storage using price-taker models and historic wholesale price data. Such studies generally focus on the arbitrage value of storage and, by design, cannot say much about the long-run value of storage in settings with higher renewable penetration. \cite{Sioshansi_2009} explore the value of electricity storage using historical price data from the PJM market in the USA and find a generally increasing arbitrage value between 2002 and 2007, but also large regional differences. \cite{Bradbury_2014} use historic market prices of seven US markets to optimize the sizing and dispatch of various energy storage technologies. They find very low E/P ratios for most storage technologies, and up to eight hours for pumped hydro and compressed air storage. \cite{McConnell_2015} analyze the value of electricity storage in the Australian National Energy Market and find that it hardly increases any more for E/P ratios above six hours.

\subsubsection{Time series-based models}
Analyses based on time series of renewable availability and load allow a more forward-looking investigation of the role of electricity storage for renewable integration. \cite{Heide_2010} investigate the seasonal balancing of wind and solar in a pan-European 100\% renewable scenario, under the assumption of substantial transmission infrastructure being in place. Seasonal long-term storage with a capacity equalling $1.5$–$1.8$ times the average monthly load is needed to balance excess PV generation in summer with excess wind generation in winter. In a comparable setting, \cite{Rasmussen_2012} find that a mix of short-term storage and long-term power-to-gas-to-power storage with an energy capacity of less than 1\% of the yearly load suffices for achieving a fully renewable energy supply in Europe.

For Germany, \cite{Weitemeyer_2015} show that electricity storage may not be required up to a VRE share of 50\%. Above that, storage needs increase, which is especially true for long-term storage in settings beyond 80\% renewable penetration. Also using a time series-based approach, \cite{Raynaud_2018} analyze the occurrence of energy droughts from variable renewables in Europe, i.e.,~periods of high residual load. They find that relatively small electricity storage capacity suffices to mitigate such energy draughts.

\subsubsection{Capacity expansion models}

In another strand of literature, detailed power sector models are used to analyze the role of electricity storage in future VRE scenarios. In general, such models are able to consider both the arbitrage and capacity values of storage. Some models further allow covering additional system values, such as the contribution of storage to the provision of balancing reserves.

The literature agrees that storage power capacity remains moderate even in scenarios with substantial carbon emissions reductions. However, the need for longer-term storage significantly increases as VRE penetration approaches 100\% and in the absence of other flexibility options. \cite{Schill_2018} show this for Germany, \cite{Safaei_2015} and \cite{deSisternes_2016} for Texas, \cite{Tong_2020} for the contiguous USA, and \cite{Ziegler_2019} for four US regions. \cite{Dowling_2020} likewise highlight the importance of long-duration electricity storage for fully VRE-based power sectors.

The ``system value'' of storage, defined as the reduction of the total system costs caused by its incremental deployment, decreases with increasing storage capacity deployment \citep{Heuberger_2017}. \cite{Mallapragada_2020} show that the value of electricity storage is largely related to the deferral of generation and transmission capacity in the long run. They also find that variable renewables and storage are complements, as the penetration of one increases the value of the other, and at the same time competitors, as storage can be replaced by VRE overcapacity and curtailment. That means that although both VRE and storage values decrease as their penetration increase, ceteris paribus, their combined deployment mitigates this decline in their system value.

\subsubsection{Interaction of electricity storage and renewable curtailment}

Many analyses, including several of the ones above, deal with the interaction of renewable curtailment and electricity storage. In an early article, \cite{Denholm_2011} investigate this for future scenarios of Texas. With medium penetration of variable renewable energy sources, only small storage capacities are needed to keep renewable curtailment low; yet if the VRE share increases to 80\%, more substantial electricity storage with a capacity of about one day of average demand is required to keep renewable curtailment below 10\%. 

Using a simple dispatch model, \cite{Schill_2014} derives similar findings for mid- to long-term scenarios for Germany. He shows that electricity storage needs substantially decrease if small levels of VRE curtailment are allowed. If, on the contrary, renewable curtailment is restricted, electricity storage needs vastly increase. This is because the most extreme renewable surplus events come with very high power and energy, which then determine storage dimensioning. Similarly, avoiding must-run of thermal generators also substantially mitigates electricity storage needs. 

\cite{Sinn_2017} revisited this issue, a priori ruling out the possibility of renewable curtailment. Using a German time series of wind, solar and load, he illustrates that storage needs in such a setting would very substantially increase with increasing VRE curtailment, and argues that this may put the viability of the renewable energy transition into question. In an open-source replication study, \cite{Zerrahn_2018} show that storage needs strongly decrease if renewable curtailment is not artificially constrained. With a stylized optimization model, they further show that a mix of VRE curtailment and storage deployment minimizes overall system costs, and that optimal electricity storage capacities may decrease further if flexible Power-to-X options are also considered.

\subsection{Graphical intuition on the changing role and value of electricity storage}\label{sub: intuition storage}

In the following, we aim to provide some graphical intuition on the changing role of electricity storage under increasing VRE shares. We do so with residual load duration curves, price duration curves, and market values calculated with a stylized open-source capacity expansion model. This model is a much-reduced version of the model DIETER, which was first presented by \cite{Zerrahn_2017} and subsequently expanded and applied in various other papers, as listed in \cite{Gaete_2020}. The reduced model version, which has been used for high-level illustrations before \citep{Zerrahn_2018, Schill_Joule_2020}, determines cost-minimal generation and storage capacities as well as their optimal dispatch for different shares of VRE in final yearly demand, which is enforced by a respective constraint. This can be interpreted as a setting in which renewable generators receive an energy-based premium on top of their wholesale market revenues, such that the desired VRE share is perfectly met. The model only considers five generation technologies (hard coal, open- and  closed-cycle gas turbines, solar PV, and wind power), one stylized storage technology (parameterized as pumped hydro storage), and a single region (Germany, one node). Hourly electricity demand is assumed to be fully price-inelastic. The model is freely available under a permissive license \citep{Schill_zenodo_2020}. Given the stylized nature of this model, the focus is not on the numbers, but on qualitative insights and the general mechanisms at play. 

For a description of the full DIETER model's general setup see \cite{Zerrahn_2017}. For exemplary applications see \cite{Schill_2017}, \cite{Schill_2018}, or \cite{Say_2020}. Here we use a stylized version that builds on \cite{Zerrahn_2018} and \cite{Schill_Joule_2020}. It is a linear model that minimizes overall electricity sector costs, consisting of investment and variable costs. Its results reflect a frictionless market in a long-run equilibrium. The model is solved for all consecutive hours of a full year, using GAMS / CPLEX. The optimization is subject to a number of constraints, such as generation capacity and storage restrictions as well as an energy balance. A restriction forces the model to exactly cover a predefined share of overall yearly electricity demand by variable renewable energy sources. Note that the VRE shares reported in the paper refer to the share in overall final demand, given that storage losses are completely covered by VRE. For more details, see the model code.

Endogenous model outputs include overall costs, capacity deployment and dispatch of all generation and electricity storage technologies, and the dual variable of the energy balance, which we interpret as wholesale prices. Exogenous input data include time series of electricity demand and renewable availability, which are taken from the Open Power System Data platform \citep{Wiese2019}, as well as fixed and variable costs and efficiencies of generation and storage technologies. All inputs are loosely calibrated to a German mid-term scenario, using a greenfield approach. Input data is freely available together with the open-source code in the Zenodo repository \url{https://doi.org/10.5281/zenodo.4383288}.

\subsubsection{Residual load and price duration curves}

\begin{figure}[t]
    \centering{} \includegraphics[width=.8\textwidth]{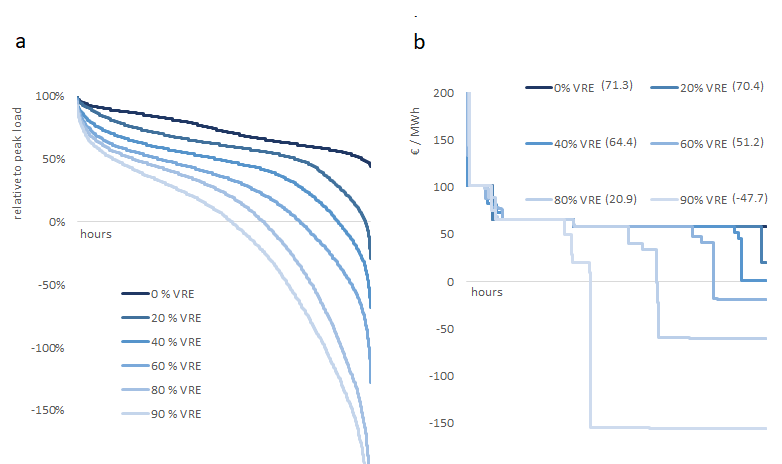}\\
    \caption{\label{fig: RLDC lines}Residual load duration curves and price duration curves for different VRE shares in a stylized German setting.}
\end{figure}

Panel a of Figure \ref{fig: RLDC lines} illustrates how the residual load duration curve (RLDC) changes with increasing shares of variable renewable energy sources \citep[cp.][]{Ueckerdt_2015}. As VRE penetration increases, the RLDC shifts downwards on its right-hand side. The number of hours during which the residual load is negative, i.e.,~the VRE generation potential exceeds the electric load, increases substantially beyond a renewable share of 20\%, and the magnitude of peak negative residual load also increases strongly \citep[cp.~also][]{Schill_2014}. The left-hand side of the RLDC hardly decreases with increasing renewable penetration, which reflects the small contribution of wind power and solar PV to the provision of firm capacity. If the VRE share increases to 90\%, renewable surplus energy occurs in nearly every second hour of the year. The shape of the RLDC also changes, which is driven by different cost-minimizing shares of wind and solar power for different VRE penetrations. In the parameterization used here, only PV is deployed in the 20\% setting, which regularly leads to low residual load values during summer days and explains the bend on the right-hand side. In settings with higher VRE shares, wind power with smoother generation profiles is increasingly deployed.

\begin{figure}[ht]
    \centering{} \includegraphics[width=.8\textwidth]{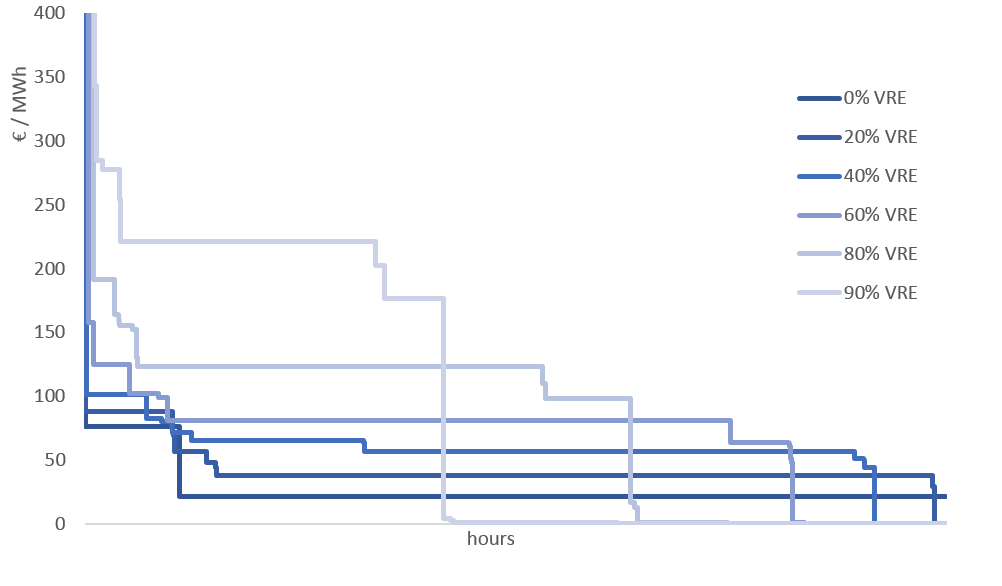}\\
    \caption{\label{fig: PDC CO2-driven}Price duration curves for different VRE shares in a stylized German setting.}
\end{figure}

Panel b of Figure \ref{fig: RLDC lines} shows price duration curves (PDCs) for the same illustrative example, as well as weighted average wholesale prices. The ordinate is capped, leaving out high peak prices in a few hours to improve visibility. Using a cost-minimization approach, we do not explicitly model market prices, but interpret the dual variable of the energy constraint as a wholesale price, as often done in model-based energy economic analyses \citep{Brown_2020}. With increasing VRE penetration, the PDC shifts downward on the right-hand side, similarly to the RLDC, which corresponds to higher renewable surplus generation. Prices may even become negative, which reflects the fact that the objective value, i.e.,~overall system costs, would decrease if more electricity would be consumed in this hour. This is because the binding VRE constraint could be easier achieved if consumption increased during hours of renewable curtailment. In an alternative setting, where VRE deployment is not driven by a constraint (i.e.,~by energy-based support), but by \coo pricing, wholesale prices never turn negative (Figure~\ref{fig: PDC CO2-driven}). When interpreting this results, the reader should also keep in mind that electricity consumers usually also have to carry the costs of renewable support, so electricity bills never become negative. We further assume a fully price-inelastic demand, and that the total system costs only partially include the external costs of conventional generators.

\subsubsection{The changing role of storage}

\begin{figure}[t]
    \centering{} \includegraphics[width=1.0\textwidth]{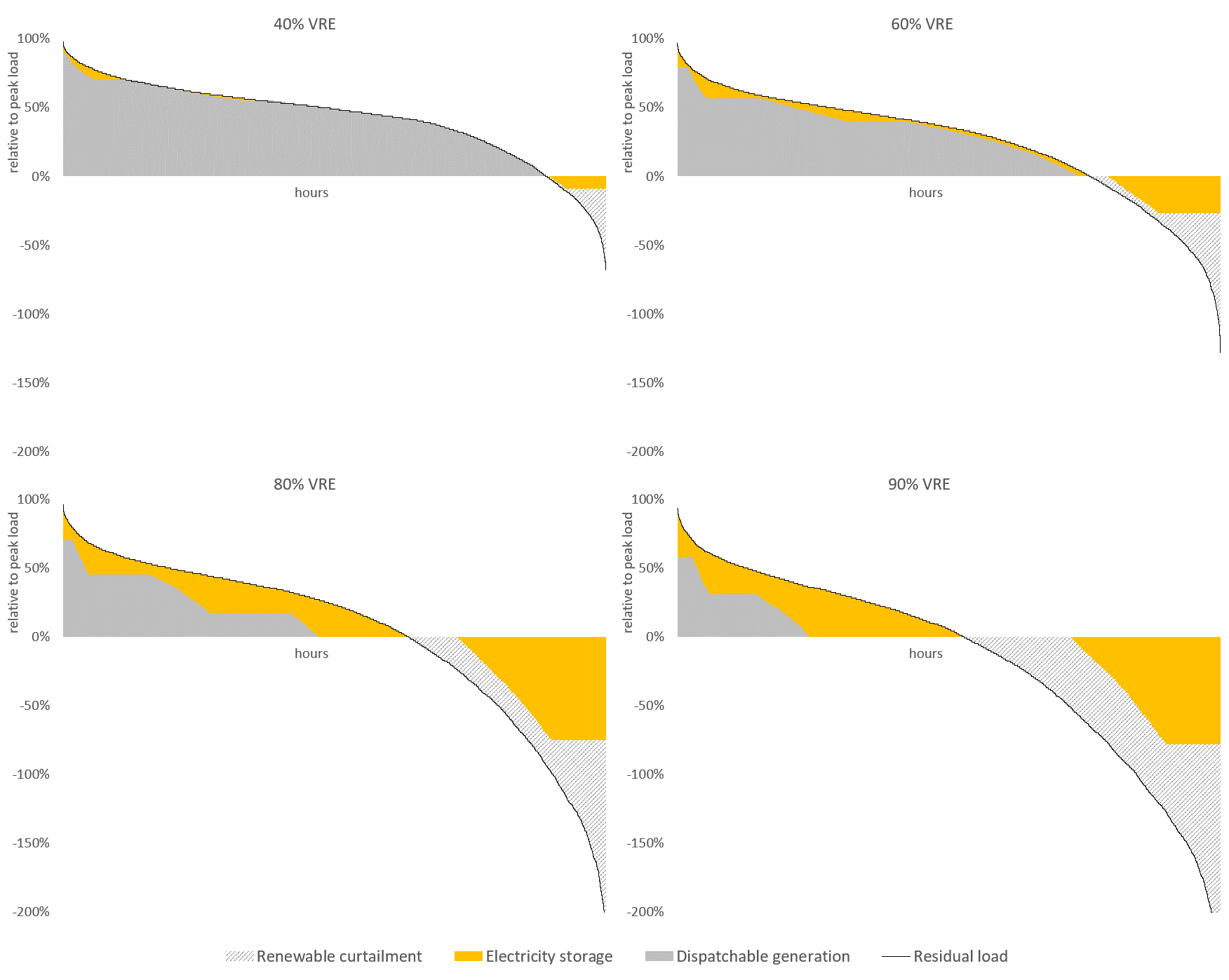}\\
    \caption{\label{fig: RLDC storage}Residual load and optimal electricity storage use for different VRE shares in a stylized German setting.}
\end{figure}

Figure \ref{fig: RLDC storage} shows the changing use of electricity storage with increasing VRE penetration. In the stylized setting used here, no storage is deployed below a VRE share of 30\%. For medium VRE shares, optimal storage investments are moderate, but they increase disproportionately with higher variable renewable penetration. For 40\% VRE, the optimal storage power rating corresponds to 9\% of the annual peak load. If the VRE share increases to 60\%, 80\% or 90\%, storage power increases to 27\%, 75\% and 78\% of the peak load. The optimal storage energy capacity increases even more, from 0.01\% of the yearly electricity demand in the 40\% VRE case to 0.03\% (60\%), 0.11\% (80\%), and 0.21\% (90\%).

Accordingly, storage is increasingly used with higher VRE penetration, as shown in Figure~\ref{fig: RLDC storage}. Importantly, storage is never deployed to fully take up renewable surplus generation in a least-cost solution, as this would require excessive and under-utilized investments into storage power and, even more so, storage energy capacity \citep[cp.][]{Schill_2014, Zerrahn_2018}. In a cost-minimizing solution, there will accordingly always be some level of renewable curtailment, absent geographical balancing, flexible Power-to-X, or other low-cost flexibility options \citep[cp.][]{Schill_Joule_2020}. Figure~\ref{fig: RLDC storage} also shows that storage is used to smooth the generation of the three dispatchable technologies modeled here, i.e.,~to optimize their full-load hours, depending on their fixed and variable cost structure (cp.~section~\ref{sub: dispatchable vs. variable}). The use of hard coal, which is the ``baseload'' technology in this setting, decreases with increasing renewable penetration and is completely crowded out in the 90\% VRE case.

Electricity storage hardly contributes to peak residual load supply in settings with moderate VRE shares. Instead, storage deployment is driven mainly by taking up renewable surplus energy in the 40\% and 60\% VRE cases, i.e.,~by the right-hand side of the RLDC. This changes for very high VRE shares, where storage is increasingly used to also supply peak residual load on the right-hand side of the RLDC. This also requires larger investments in storage energy capacity, as increasing amounts of renewable surplus energy have to be shifted over longer periods of time. In 100\% VRE settings, where no other X-to-Power options are left, electricity storage would have to supply the total peak residual load and fully balance seasonal fluctuations.

\subsubsection{Wind and solar cost and value}

\begin{figure}[t]
    \centering{} \includegraphics[width=.8\textwidth]{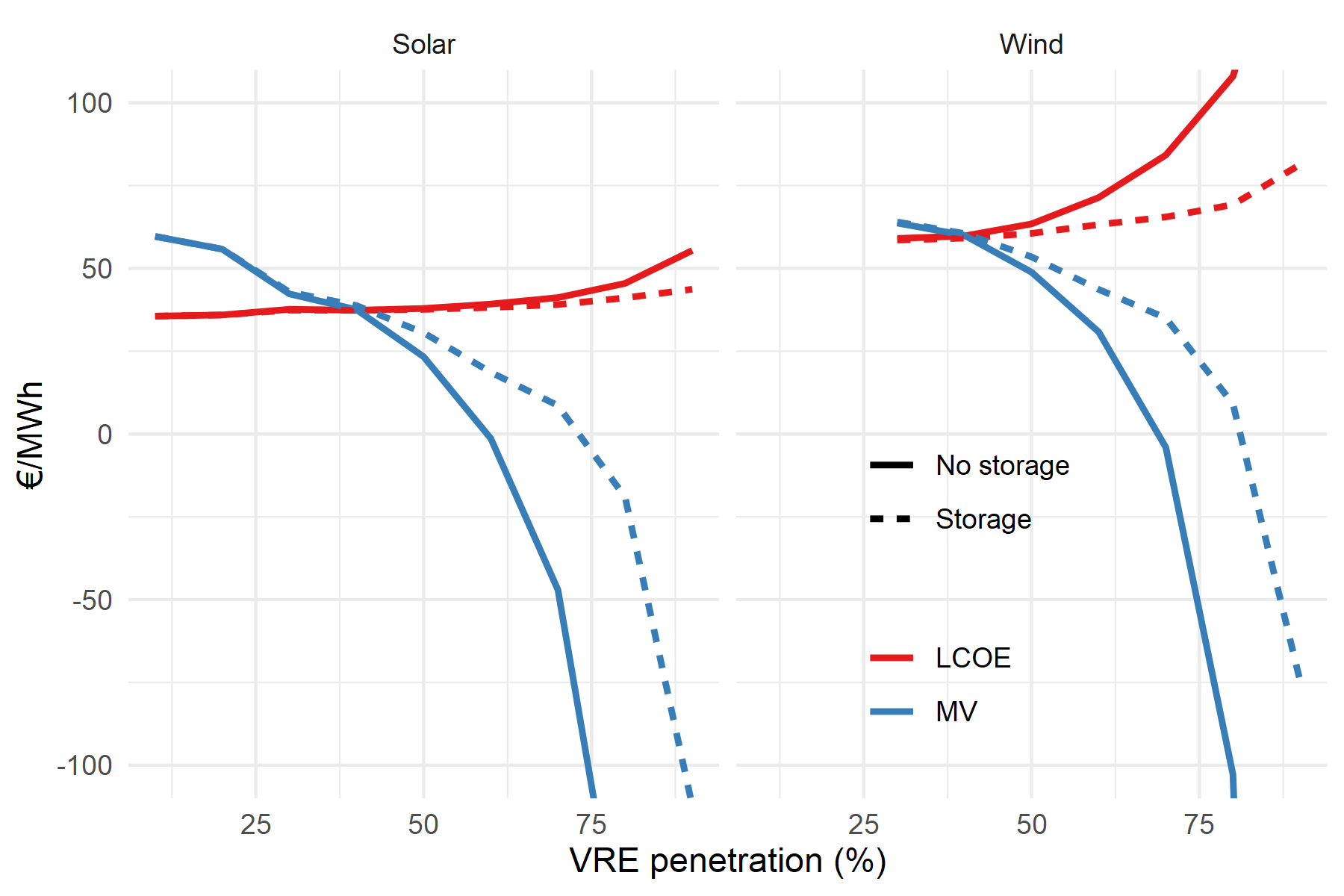}\\
    \caption{\label{fig: MV}Market values for wind power and solar PV and levelized costs of electricity for increasing VRE shares in a stylized German setting.}
\end{figure}

Figure \ref{fig: MV} illustrates the market values and LCOE for wind power and solar PV for the same stylized example as above and also in an equivalent setting without storage. The cannibalization effect is clearly visible: with increasing VRE penetration, the market values of wind and solar deteriorate. This is particularly true for solar PV, driven by regular diurnal generation patterns. This finding is in line with the empirical literature discussed in section~\ref{sub: cannibalization effect}, as for example shown by \cite{LopezProl2020}. 

\begin{figure}[ht]
    \centering{} \includegraphics[width=.8\textwidth]{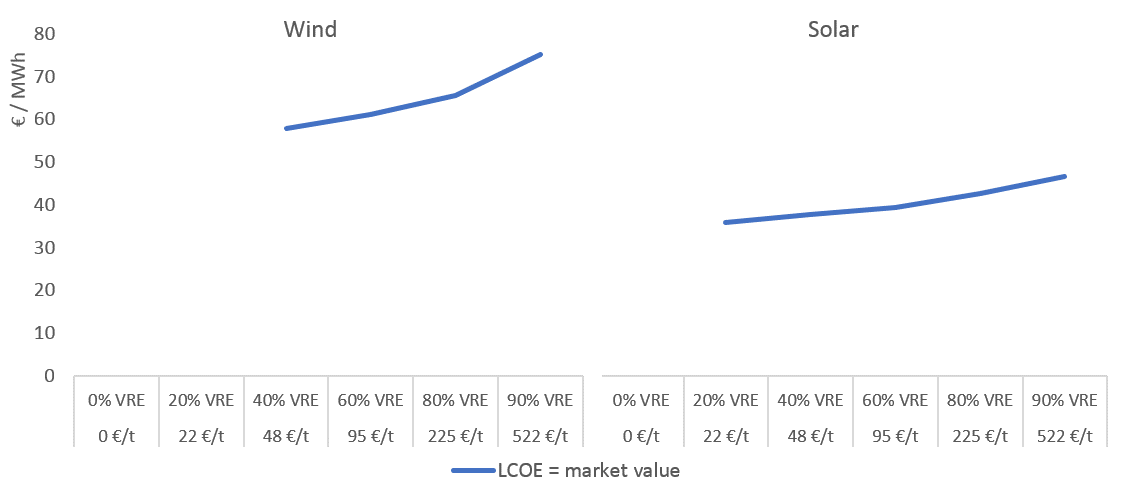}\\
    \caption{\label{fig: MV CO2-price-driven}Market values for wind power and solar PV and levelized costs of electricity for increasing VRE shares driven by \coo pricing in a stylized German setting.}
\end{figure}

At the same time, the levelized costs of wind and solar energy increase, reflecting increasing levels of curtailment. The difference between the market value and the LCOE can be interpreted as the required renewable support per unit, which also equals the dual of the VRE constraint in the optimization model. Note that up to 40\%~VRE, the market value of wind an solar is greater than their LCOE, which reflects the fact that the cost-minimizing renewable share (only partially considering the externalities of conventional generators) in this stylized parameterization is somewhat above 40\%. It should be noted that the decreasing market value is driven by negative prices caused by the binding VRE constraint in the model. In an alternative model setting, in which increasing VRE shares are not driven by a respective constraint, but by increasing \coo pricing, wholesale prices never become negative. Likewise, the market values of wind and solar do not deteriorate in such a setting with increasing VRE shares, but always correspond to the respective levelized costs. This is shown by \cite{Brown_2020}, and we also illustrate this in Figure~\ref{fig: MV CO2-price-driven}. This is because increasing \coo prices internalizes climate change costs and therefore allow VRE to capture that mitigation value, compensating thus the cannibalization effect of their increasing penetration.

\begin{figure}[ht]
    \centering
    \includegraphics[width=12cm]{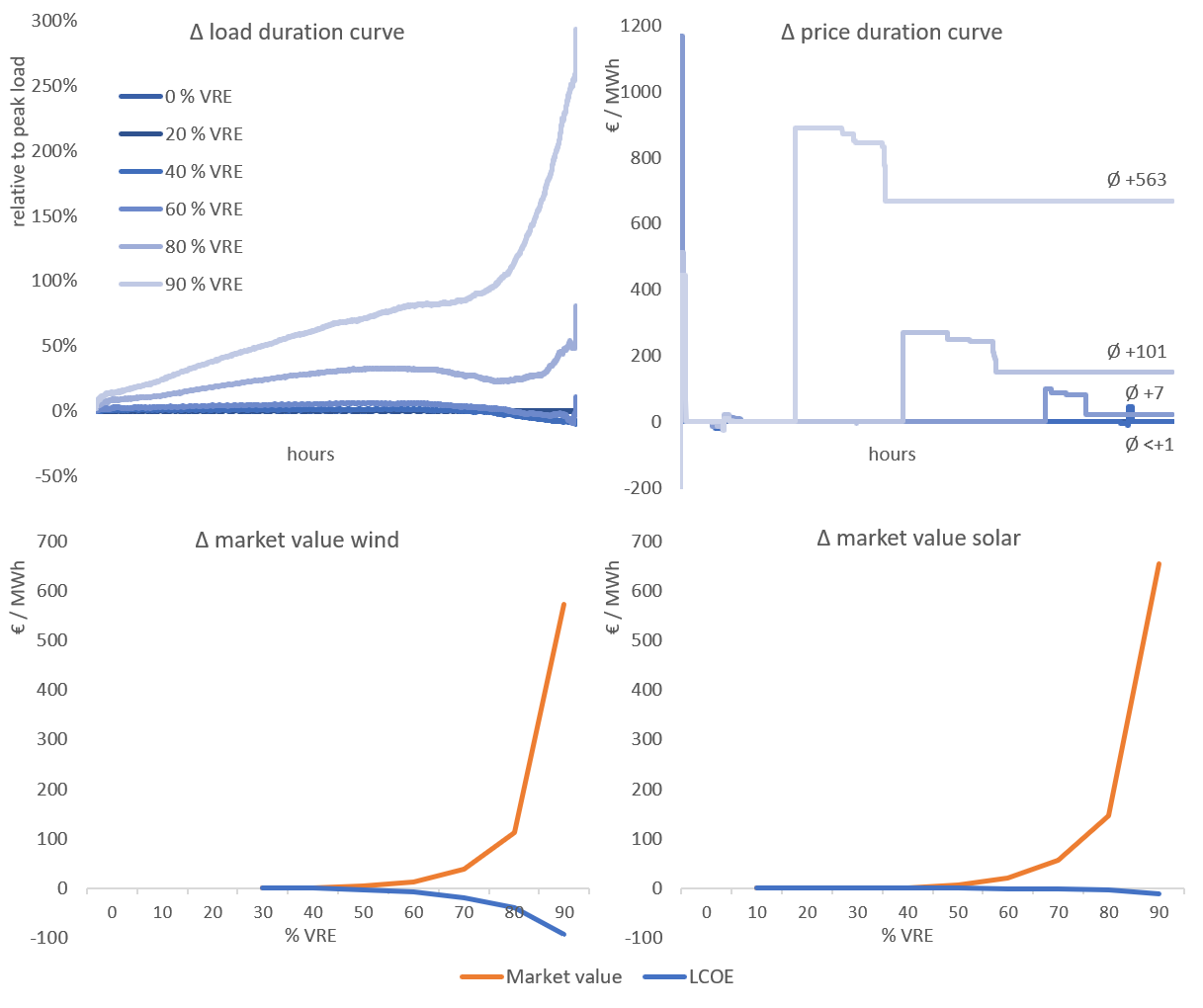}
    \caption{Differences to a setting without storage: changes of the load and price duration curves, as well as changes in the market values and LCOE of wind power and solar PV in a stylized German setting.}
    \label{fig: delta storage}
\end{figure}

The comparison between the market values of wind and solar with and without storage reveals the effect of storage on mitigating the cannibalization effect. Figure~\ref{fig: delta storage} illustrates the effects of storage on residual load and price duration curves and VRE market values in more detail. The decline of VRE unit revenues, particularly for solar, would be stronger in the absence of storage. Likewise, their levelized cost would increase faster as penetration increases due to higher curtailment.

\subsection{Beyond stationary grid-scale electricity storage}

While we have focused on stationary grid-scale electricity storage, decentralized small-scale electricity storage is becoming more and more relevant in many markets. Decentralized PV-battery systems allow households and commercial electricity consumers to increase their levels of PV self-consumption. This can be beneficial if the self-generated electricity is cheaper than the retail tariff of the grid electricity it substitutes \citep{Schill_2017}, as it is already the case in many countries \citep{Lang2016a, LopezProl2020a}. Yet the power sector implications of widespread PV-battery adoption and their interaction with grid-scale electricity storage are only partially understood so far \citep{Say_2020}.

Battery electric vehicles are a particularly interesting case, as they entail new and potentially flexible loads in the power sector that could also feed back electricity to the grid. The latter concept has been discussed for many years under the label Vehicle-to-Grid \citep{Kempton_2005}. Accordingly, battery-electric vehicles potentially combine Power-to-X and Power-to-Power properties. Due to technical, economic and social barriers, Vehicle-to-Grid is hardly used in practice so far. Yet this may change in the near future, considering the current massive roll-out of battery-electric vehicles in many countries.

In general, various future sector coupling options could supply additional flexibility to the electricity system. For example, \cite{Brown_2018} find that flexible charging of electric vehicle batteries could balance daily PV fluctuations, and Power-to-Heat applications with thermal energy storage could balance longer-duration wind power variability. Likewise, \cite{Schill_Joule_2020} illustrates that flexible sector coupling may substantially reduce the need for electricity storage in settings with medium VRE penetration, as it reduces renewable surplus energy on the right-hand side of the RLDC. Yet in 100\%~VRE settings, this storage-mitigating effect of flexible Power-to-X options vanishes, as electricity storage is still required to supply positive residual load on the left-hand side of the RLDC. Temporally inflexible sector coupling, in contrast, would increase the flexibility needs of the power sector.

\section{Conclusions}

We have discussed the literature on the interactions between increasing variable renewable energy penetration, electricity storage, and electricity prices. Increasing penetration of VRE lowers wholesale electricity prices due to the merit-order effect. Since this happens at times of high VRE penetration, VRE market values fall (absolute cannibalization effect), and they fall even faster than electricity prices, causing a decline of technology-specific value factors (unit revenues divided by average wholesale electricity prices). The decreasing value factor (relative cannibalization effect) can be interpreted as the evolution of VRE integration costs from a value perspective. 

Although VRE penetration is generally thought to increase price volatility, empirical evidence is mixed. Effects of VRE on price volatility depend on the generation pattern of the VRE technology, the considered time scale, and the characteristics of the electricity system itself, such as availability of flexibility, demand patterns and market design.

Electricity storage and other flexibility options mitigate the cannibalization effect of VRE and stabilize electricity prices. VRE and electricity storage are complementary in the sense that higher penetration of one increases the value of the other. However, they are to some degree also substitutes as storage can be replaced by VRE overcapacity and curtailment. Storage also shows diminishing marginal returns as each additional unit of capacity provides lower value to the system.

Whereas we have focused here on electricity storage, several other forms of power sector flexibility can contribute to integrating high shares of variable renewables. This includes Power-to-Power options such as demand response, Power-to-X options related to sector coupling, as well as geographical balancing. Both the optimal mix of such flexibility options and their market interactions are not fully understood so far.

Future research should thus address in more detail the interactions between grid-scale and decentralized electricity storage, as well as various sector coupling options. This also requires taking the operational characteristics of decentralized batteries into account, which heavily depend on end-user behaviour as well as on retail market design. Likewise, the flexibility potentials and constraints of various Power-to-X processes and their repercussions on electricity storage and VRE market dynamics should be studied in more detail. 

Further, market design and regulatory frameworks may have to be adjusted in order to enable all flexibility options to capture the various values they may provide to the system. This is necessary to achieve welfare-optimal deployment of each technology in the real world. Yet the necessary adjustments may differ strongly between markets and jurisdictions, and are also subject to further research.


\section*{Acknowledgments}
Javier López Prol acknowledges funding of the Austrian Science Fund (FWF) via the Schrödinger Fellowship ``Cannibalization'' (grant number J 4301-G27). Wolf-Peter Schill acknowledges research grants by the German Federal Ministry of Education and Research via the project ``Future of Fossil Fuels in the wake of greenhouse gas neutrality'' (grant number FKZ 01LA1810B), and by the German Federal Ministry for Economic Affairs and Energy via the project ``MODEZEEN'' (grant number FKZ 03EI1019D).

\bibliography{manuscript}

\begin{thebibliography}{133}
\expandafter\ifx\csname natexlab\endcsname\relax\def\natexlab#1{#1}\fi
\providecommand{\url}[1]{\texttt{#1}}
\providecommand{\href}[2]{#2}
\providecommand{\path}[1]{#1}
\providecommand{\DOIprefix}{doi:}
\providecommand{\ArXivprefix}{arXiv:}
\providecommand{\URLprefix}{URL: }
\providecommand{\Pubmedprefix}{pmid:}
\providecommand{\doi}[1]{\href{http://dx.doi.org/#1}{\path{#1}}}
\providecommand{\Pubmed}[1]{\href{pmid:#1}{\path{#1}}}
\providecommand{\bibinfo}[2]{#2}
\ifx\xfnm\relax \def\xfnm[#1]{\unskip,\space#1}\fi
\bibitem[{Aghahosseini et~al.(2019)Aghahosseini, Bogdanov, Barbosa and
  Breyer}]{Aghahosseini2019}
\bibinfo{author}{Aghahosseini, A.}, \bibinfo{author}{Bogdanov, D.},
  \bibinfo{author}{Barbosa, L.S.}, \bibinfo{author}{Breyer, C.},
  \bibinfo{year}{2019}.
\newblock \bibinfo{title}{{Analysing the feasibility of powering the Americas
  with renewable energy and inter-regional grid interconnections by 2030}}.
\newblock \bibinfo{journal}{Renew. Sustain. Energy Rev.} \bibinfo{volume}{105},
  \bibinfo{pages}{187--205}.
\newblock \DOIprefix\doi{10.1016/j.rser.2019.01.046}.
\bibitem[{{Agora Energiewende}(2017)}]{Agora2017}
\bibinfo{author}{{Agora Energiewende}}, \bibinfo{year}{2017}.
\newblock \bibinfo{title}{{Future Cost of Onshore Wind. Recent auction results,
  long-term outlook and implications for upcoming German auctions}}.
\newblock \bibinfo{type}{Technical Report}.
\newblock \URLprefix
  \url{https://www.agora-energiewende.de/fileadmin2/Projekte/2017/Future_Cost_of_Wind/Agora_Future-Cost-of-Wind_WEB.pdf}.
\bibitem[{Armaroli and Balzani(2011)}]{Armaroli_2011}
\bibinfo{author}{Armaroli, N.}, \bibinfo{author}{Balzani, V.},
  \bibinfo{year}{2011}.
\newblock \bibinfo{title}{Towards an electricity-powered world}.
\newblock \bibinfo{journal}{Energy \& Environmental Science}
  \bibinfo{volume}{4}, \bibinfo{pages}{3193--3222}.
\newblock \DOIprefix\doi{10.1039/C1EE01249E}.
\bibitem[{Baker et~al.(2013)Baker, Fowlie, Lemoine and Reynolds}]{Baker2013}
\bibinfo{author}{Baker, E.}, \bibinfo{author}{Fowlie, M.},
  \bibinfo{author}{Lemoine, D.}, \bibinfo{author}{Reynolds, S.S.},
  \bibinfo{year}{2013}.
\newblock \bibinfo{title}{{The Economics of Solar Electricity}}.
\newblock \bibinfo{journal}{Annu. Rev. Resour. Econ.} \bibinfo{volume}{5},
  \bibinfo{pages}{387--426}.
\newblock \DOIprefix\doi{10.1146/annurev-resource-091912-151843}.
\bibitem[{Ballester and Furi{\'{o}}(2015)}]{Ballester2015}
\bibinfo{author}{Ballester, C.}, \bibinfo{author}{Furi{\'{o}}, D.},
  \bibinfo{year}{2015}.
\newblock \bibinfo{title}{{Effects of renewables on the stylized facts of
  electricity prices}}.
\newblock \bibinfo{journal}{Renewable and Sustainable Energy Reviews}
  \bibinfo{volume}{52}, \bibinfo{pages}{1596--1609}.
\newblock \DOIprefix\doi{10.1016/j.rser.2015.07.168}.
\bibitem[{Beuse et~al.(2020)Beuse, Steffen and Schmidt}]{Beuse2020}
\bibinfo{author}{Beuse, M.}, \bibinfo{author}{Steffen, B.},
  \bibinfo{author}{Schmidt, T.S.}, \bibinfo{year}{2020}.
\newblock \bibinfo{title}{{Projecting the Competition between Energy-Storage
  Technologies in the Electricity Sector}}.
\newblock \bibinfo{journal}{Joule} \bibinfo{volume}{4},
  \bibinfo{pages}{2162--2184}.
\newblock \DOIprefix\doi{10.1016/j.joule.2020.07.017}.
\bibitem[{Bistline et~al.(2020)Bistline, Cole, Damato, DeCarolis, Frazier,
  Linga, Marcy, Namovicz, Podkaminer, Sims, Sukunta and Young}]{Bistline_2020}
\bibinfo{author}{Bistline, J.}, \bibinfo{author}{Cole, W.},
  \bibinfo{author}{Damato, G.}, \bibinfo{author}{DeCarolis, J.},
  \bibinfo{author}{Frazier, W.}, \bibinfo{author}{Linga, V.},
  \bibinfo{author}{Marcy, C.}, \bibinfo{author}{Namovicz, C.},
  \bibinfo{author}{Podkaminer, K.}, \bibinfo{author}{Sims, R.},
  \bibinfo{author}{Sukunta, M.}, \bibinfo{author}{Young, D.},
  \bibinfo{year}{2020}.
\newblock \bibinfo{title}{Energy storage in long-term system models: a review
  of considerations, best practices, and research needs}.
\newblock \bibinfo{journal}{Progress in Energy} \bibinfo{volume}{2},
  \bibinfo{pages}{032001}.
\newblock \DOIprefix\doi{10.1088/2516-1083/ab9894}.
\bibitem[{Blanco and Faaij(2018)}]{Blanco_2018}
\bibinfo{author}{Blanco, H.}, \bibinfo{author}{Faaij, A.},
  \bibinfo{year}{2018}.
\newblock \bibinfo{title}{A review at the role of storage in energy systems
  with a focus on {Power to Gas} and long-term storage}.
\newblock \bibinfo{journal}{Renewable and Sustainable Energy Reviews}
  \bibinfo{volume}{81}, \bibinfo{pages}{1049 -- 1086}.
\newblock \DOIprefix\doi{10.1016/j.rser.2017.07.062}.
\bibitem[{Bloess et~al.(2018)Bloess, Schill and Zerrahn}]{Bloess_2018}
\bibinfo{author}{Bloess, A.}, \bibinfo{author}{Schill, W.P.},
  \bibinfo{author}{Zerrahn, A.}, \bibinfo{year}{2018}.
\newblock \bibinfo{title}{Power-to-heat for renewable energy integration: A
  review of technologies, modeling approaches, and flexibility potentials}.
\newblock \bibinfo{journal}{Applied Energy} \bibinfo{volume}{212},
  \bibinfo{pages}{1611 -- 1626}.
\newblock \DOIprefix\doi{10.1016/j.apenergy.2017.12.073}.
\bibitem[{Boiteux(1949)}]{Boiteux1949}
\bibinfo{author}{Boiteux, M.}, \bibinfo{year}{1949}.
\newblock \bibinfo{title}{{Peak-Load Pricing}}.
\newblock \bibinfo{journal}{J. Bus.} \bibinfo{volume}{33},
  \bibinfo{pages}{157--179}.
\newblock \DOIprefix\doi{10.2307/2351015}.
\bibitem[{{BP}(2020)}]{BP2020}
\bibinfo{author}{{BP}}, \bibinfo{year}{2020}.
\newblock \bibinfo{title}{{BP Statistical Review of World Energy 2020}}.
\newblock \bibinfo{type}{Technical Report}.
\newblock \URLprefix
  \url{https://www.bp.com/content/dam/bp/business-sites/en/global/corporate/pdfs/energy-economics/statistical-review/bp-stats-review-2020-full-report.pdf}.
\bibitem[{Bradbury et~al.(2014)Bradbury, Pratson and
  Patino-Echeverri}]{Bradbury_2014}
\bibinfo{author}{Bradbury, K.}, \bibinfo{author}{Pratson, L.},
  \bibinfo{author}{Patino-Echeverri, D.}, \bibinfo{year}{2014}.
\newblock \bibinfo{title}{Economic viability of energy storage systems based on
  price arbitrage potential in real-time {U.S.} electricity markets}.
\newblock \bibinfo{journal}{Applied Energy} \bibinfo{volume}{114},
  \bibinfo{pages}{512 -- 519}.
\newblock \DOIprefix\doi{10.1016/j.apenergy.2013.10.010}.
\bibitem[{Brown et~al.(2018a)Brown, Bischof-Niemz, Blok, Breyer, Lund and
  Mathiesen}]{Brown_response_2018}
\bibinfo{author}{Brown, T.}, \bibinfo{author}{Bischof-Niemz, T.},
  \bibinfo{author}{Blok, K.}, \bibinfo{author}{Breyer, C.},
  \bibinfo{author}{Lund, H.}, \bibinfo{author}{Mathiesen, B.},
  \bibinfo{year}{2018}a.
\newblock \bibinfo{title}{Response to 'burden of proof: A comprehensive review
  of the feasibility of 100\% renewable-electricity systems'}.
\newblock \bibinfo{journal}{Renewable and Sustainable Energy Reviews}
  \bibinfo{volume}{92}, \bibinfo{pages}{834 -- 847}.
\newblock \DOIprefix\doi{10.1016/j.rser.2018.04.113}.
\bibitem[{Brown and Reichenberg(2020)}]{Brown_2020}
\bibinfo{author}{Brown, T.}, \bibinfo{author}{Reichenberg, L.},
  \bibinfo{year}{2020}.
\newblock \bibinfo{title}{Decreasing market value of variable renewables is a
  result of policy, not variability}.
\newblock \href{http://arxiv.org/abs/2002.05209}{{\tt arXiv:2002.05209}}.
\bibitem[{Brown et~al.(2018b)Brown, Schlachtberger, Kies, Schramm and
  Greiner}]{Brown_2018}
\bibinfo{author}{Brown, T.}, \bibinfo{author}{Schlachtberger, D.},
  \bibinfo{author}{Kies, A.}, \bibinfo{author}{Schramm, S.},
  \bibinfo{author}{Greiner, M.}, \bibinfo{year}{2018}b.
\newblock \bibinfo{title}{Synergies of sector coupling and transmission
  reinforcement in a cost-optimised, highly renewable {European} energy
  system}.
\newblock \bibinfo{journal}{Energy} \bibinfo{volume}{160}, \bibinfo{pages}{720
  -- 739}.
\newblock \DOIprefix\doi{10.1016/j.energy.2018.06.222}.
\bibitem[{Bushnell and Novan(2018)}]{Bushnell2018}
\bibinfo{author}{Bushnell, J.}, \bibinfo{author}{Novan, K.},
  \bibinfo{year}{2018}.
\newblock \bibinfo{title}{{Setting with the Sun: The Impacts of Renewable
  Energy on Wholesale Power Markets}}.
\newblock \bibinfo{type}{Working Paper}. National Bureau of Economic Research.
\newblock \DOIprefix\doi{10.3386/w24980}.
\bibitem[{Castillo and Gayme(2014)}]{Castillo_2014}
\bibinfo{author}{Castillo, A.}, \bibinfo{author}{Gayme, D.F.},
  \bibinfo{year}{2014}.
\newblock \bibinfo{title}{Grid-scale energy storage applications in renewable
  energy integration: A survey}.
\newblock \bibinfo{journal}{Energy Conversion and Management}
  \bibinfo{volume}{87}, \bibinfo{pages}{885 -- 894}.
\newblock \DOIprefix\doi{10.1016/j.enconman.2014.07.063}.
\bibitem[{Child et~al.(2019)Child, Kemfert, Bogdanov and Breyer}]{Child_2019}
\bibinfo{author}{Child, M.}, \bibinfo{author}{Kemfert, C.},
  \bibinfo{author}{Bogdanov, D.}, \bibinfo{author}{Breyer, C.},
  \bibinfo{year}{2019}.
\newblock \bibinfo{title}{Flexible electricity generation, grid exchange and
  storage for the transition to a 100\% renewable energy system in {Europe}}.
\newblock \bibinfo{journal}{Renewable Energy} \bibinfo{volume}{139},
  \bibinfo{pages}{80 -- 101}.
\newblock \DOIprefix\doi{10.1016/j.renene.2019.02.077}.
\bibitem[{Ciarreta et~al.(2020)Ciarreta, Pizarro-Irizar and
  Zarraga}]{Ciarreta2020}
\bibinfo{author}{Ciarreta, A.}, \bibinfo{author}{Pizarro-Irizar, C.},
  \bibinfo{author}{Zarraga, A.}, \bibinfo{year}{2020}.
\newblock \bibinfo{title}{{Renewable energy regulation and structural breaks:
  An empirical analysis of Spanish electricity price volatility}}.
\newblock \bibinfo{journal}{Energy Econ.} \bibinfo{volume}{88},
  \bibinfo{pages}{104749}.
\newblock \DOIprefix\doi{10.1016/j.eneco.2020.104749}.
\bibitem[{Cl{\`{o}} et~al.(2015)Cl{\`{o}}, Cataldi and Zoppoli}]{Clo2015a}
\bibinfo{author}{Cl{\`{o}}, S.}, \bibinfo{author}{Cataldi, A.},
  \bibinfo{author}{Zoppoli, P.}, \bibinfo{year}{2015}.
\newblock \bibinfo{title}{{The merit-order effect in the Italian power market:
  The impact of solar and wind generation on national wholesale electricity
  prices}}.
\newblock \bibinfo{journal}{Energy Policy} \bibinfo{volume}{77},
  \bibinfo{pages}{79--88}.
\newblock \DOIprefix\doi{10.1016/j.enpol.2014.11.038}.
\bibitem[{Cl{\`{o}} and D'Adamo(2015)}]{Clo2015}
\bibinfo{author}{Cl{\`{o}}, S.}, \bibinfo{author}{D'Adamo, G.},
  \bibinfo{year}{2015}.
\newblock \bibinfo{title}{{The dark side of the sun: How solar power production
  affects the market value of solar and gas sources}}.
\newblock \bibinfo{journal}{Energy Econ.} \bibinfo{volume}{49},
  \bibinfo{pages}{523--530}.
\newblock \DOIprefix\doi{10.1016/j.eneco.2015.03.025}.
\bibitem[{Cludius et~al.(2014)Cludius, Hermann, Matthes and
  Graichen}]{Cludius2014}
\bibinfo{author}{Cludius, J.}, \bibinfo{author}{Hermann, H.},
  \bibinfo{author}{Matthes, F.C.}, \bibinfo{author}{Graichen, V.},
  \bibinfo{year}{2014}.
\newblock \bibinfo{title}{{The merit order effect of wind and photovoltaic
  electricity generation in Germany 2008-2016: Estimation and distributional
  implications}}.
\newblock \bibinfo{journal}{Energy Econ.} \bibinfo{volume}{44},
  \bibinfo{pages}{302--313}.
\newblock \DOIprefix\doi{10.1016/j.eneco.2014.04.020}.
\bibitem[{de~Coninck et~al.(2018)de~Coninck, Revi, Babiker, Bertoldi,
  Buckeridge, Cartwright, Dong, Ford, Fuss, Hourcade, Ley, Mechler, Newman,
  Revokatova, Schultz, Steg and Sugiyama}]{IPCC_2018}
\bibinfo{author}{de~Coninck, H.}, \bibinfo{author}{Revi, A.},
  \bibinfo{author}{Babiker, M.}, \bibinfo{author}{Bertoldi, P.},
  \bibinfo{author}{Buckeridge, M.}, \bibinfo{author}{Cartwright, A.},
  \bibinfo{author}{Dong, W.}, \bibinfo{author}{Ford, J.},
  \bibinfo{author}{Fuss, S.}, \bibinfo{author}{Hourcade, J.C.},
  \bibinfo{author}{Ley, D.}, \bibinfo{author}{Mechler, R.},
  \bibinfo{author}{Newman, P.}, \bibinfo{author}{Revokatova, A.},
  \bibinfo{author}{Schultz, S.}, \bibinfo{author}{Steg, L.},
  \bibinfo{author}{Sugiyama, T.}, \bibinfo{year}{2018}.
\newblock \bibinfo{title}{Strengthening and implementing the global response},
  in: \bibinfo{booktitle}{Global {Warming} of 1.5$^{\circ}${C}. {An} {IPCC}
  {Special} {Report} on the impacts of global warming of 1.5$^{\circ}${C} above
  pre-industrial levels and related global greenhouse gas emission pathways, in
  the context of strengthening the global response to the threat of climate
  change, sustainable development, and efforts to eradicate poverty}.
\newblock \URLprefix
  \url{https://www.ipcc.ch/site/assets/uploads/sites/2/2019/02/SR15_Chapter4_Low_Res.pdf}.
\bibitem[{Cramton(2017)}]{Cramton2017}
\bibinfo{author}{Cramton, P.}, \bibinfo{year}{2017}.
\newblock \bibinfo{title}{{Electricity market design}}.
\newblock \bibinfo{journal}{Oxford Rev. Econ. Policy} \bibinfo{volume}{33},
  \bibinfo{pages}{589--612}.
\newblock \DOIprefix\doi{10.1093/oxrep/grx041}.
\bibitem[{Cramton and Ockenfels(2012)}]{Cramton_2012}
\bibinfo{author}{Cramton, P.}, \bibinfo{author}{Ockenfels, A.},
  \bibinfo{year}{2012}.
\newblock \bibinfo{title}{Economics and design of capacity markets for the
  power sector}.
\newblock \bibinfo{journal}{Zeitschrift f\"{u}r Energiewirtschaft}
  \bibinfo{volume}{36}, \bibinfo{pages}{113--134}.
\newblock \DOIprefix\doi{10.1007/s12398-012-0084-2}.
\bibitem[{Creutzig et~al.(2017)Creutzig, Agoston, Goldschmidt, Luderer, Nemet
  and Pietzcker}]{Creutzig2017}
\bibinfo{author}{Creutzig, F.}, \bibinfo{author}{Agoston, P.},
  \bibinfo{author}{Goldschmidt, J.C.}, \bibinfo{author}{Luderer, G.},
  \bibinfo{author}{Nemet, G.}, \bibinfo{author}{Pietzcker, R.C.},
  \bibinfo{year}{2017}.
\newblock \bibinfo{title}{{The underestimated potential of solar energy to
  mitigate climate change}}.
\newblock \bibinfo{journal}{Nat. Energy} \bibinfo{volume}{2},
  \bibinfo{pages}{1--9}.
\newblock \DOIprefix\doi{10.1038/nenergy.2017.140}.
\bibitem[{Csereklyei et~al.(2019)Csereklyei, Qu and Ancev}]{Csereklyei2019}
\bibinfo{author}{Csereklyei, Z.}, \bibinfo{author}{Qu, S.},
  \bibinfo{author}{Ancev, T.}, \bibinfo{year}{2019}.
\newblock \bibinfo{title}{{The effect of wind and solar power generation on
  wholesale electricity prices in Australia}}.
\newblock \bibinfo{journal}{Energy Policy} \bibinfo{volume}{131},
  \bibinfo{pages}{358--369}.
\newblock \DOIprefix\doi{10.1016/j.enpol.2019.04.007}.
\bibitem[{Denholm et~al.(2010)Denholm, Ela, Kirby and Milligan}]{Denholm_2010}
\bibinfo{author}{Denholm, P.}, \bibinfo{author}{Ela, E.},
  \bibinfo{author}{Kirby, B.}, \bibinfo{author}{Milligan, M.},
  \bibinfo{year}{2010}.
\newblock \bibinfo{title}{The Role of Energy Storage with Renewable Electricity
  Generation}.
\newblock \bibinfo{type}{Technical Report} \bibinfo{number}{NREL/TP-6A2-47187}.
  National Renewable Energy Lab.(NREL), Golden, CO (United States).
\newblock \DOIprefix\doi{10.2172/972169}.
\bibitem[{Denholm and Hand(2011)}]{Denholm_2011}
\bibinfo{author}{Denholm, P.}, \bibinfo{author}{Hand, M.},
  \bibinfo{year}{2011}.
\newblock \bibinfo{title}{Grid flexibility and storage required to achieve very
  high penetration of variable renewable electricity}.
\newblock \bibinfo{journal}{Energy Policy} \bibinfo{volume}{39},
  \bibinfo{pages}{1817 -- 1830}.
\newblock \DOIprefix\doi{10.1016/j.enpol.2011.01.019}.
\bibitem[{Denholm et~al.(2013)Denholm, Jorgenson, Hummon, Palchak, Kirby, Ma
  and O'Malley}]{Denholm_2013}
\bibinfo{author}{Denholm, P.}, \bibinfo{author}{Jorgenson, J.},
  \bibinfo{author}{Hummon, M.}, \bibinfo{author}{Palchak, D.},
  \bibinfo{author}{Kirby, B.}, \bibinfo{author}{Ma, O.},
  \bibinfo{author}{O'Malley, M.}, \bibinfo{year}{2013}.
\newblock \bibinfo{title}{The Impact of Wind and Solar on the Value of Energy
  Storage}.
\newblock \bibinfo{type}{Technical Report}
  \bibinfo{number}{NREL/TP-6A20-60568}. National Renewable Energy Laboratory.
  \bibinfo{address}{15013 Denver West Parkway, Golden, CO 80401}.
\newblock \URLprefix \url{https://www.nrel.gov/docs/fy14osti/60568.pdf}.
\bibitem[{Dowling et~al.(2020)Dowling, Rinaldi, Ruggles, Davis, Yuan, Tong,
  Lewis and Caldeira}]{Dowling_2020}
\bibinfo{author}{Dowling, J.A.}, \bibinfo{author}{Rinaldi, K.Z.},
  \bibinfo{author}{Ruggles, T.H.}, \bibinfo{author}{Davis, S.J.},
  \bibinfo{author}{Yuan, M.}, \bibinfo{author}{Tong, F.},
  \bibinfo{author}{Lewis, N.S.}, \bibinfo{author}{Caldeira, K.},
  \bibinfo{year}{2020}.
\newblock \bibinfo{title}{Role of long-duration energy storage in variable
  renewable electricity systems}.
\newblock \bibinfo{journal}{Joule} \bibinfo{volume}{4},
  \bibinfo{pages}{1907--1928}.
\newblock \DOIprefix\doi{10.1016/j.joule.2020.07.007}.
\bibitem[{Ecofys(2014)}]{Ecofys2014}
\bibinfo{author}{Ecofys}, \bibinfo{year}{2014}.
\newblock \bibinfo{title}{{Subsidies and costs of EU energy}}.
\newblock \bibinfo{type}{Technical Report}.
\newblock \URLprefix
  \url{https://ec.europa.eu/energy/sites/ener/files/documents/ECOFYS%202014%20Subsidies%20and%20costs%20of%20EU%20energy_11_Nov.pdf}.
\bibitem[{Edenhofer et~al.(2014)Edenhofer, Sokona, Minx, Farahani, Kadner,
  Seyboth, Adler, Baum, Brunner, Kriemann, {Savolainen Web Manager Steffen
  Schl{\"{o}}mer}, von Stechow and {Zwickel Senior Scientist}}]{IPCC2014}
\bibinfo{author}{Edenhofer, O.}, \bibinfo{author}{Sokona, Y.},
  \bibinfo{author}{Minx, J.C.}, \bibinfo{author}{Farahani, E.},
  \bibinfo{author}{Kadner, S.}, \bibinfo{author}{Seyboth, K.},
  \bibinfo{author}{Adler, A.}, \bibinfo{author}{Baum, I.},
  \bibinfo{author}{Brunner, S.}, \bibinfo{author}{Kriemann, B.},
  \bibinfo{author}{{Savolainen Web Manager Steffen Schl{\"{o}}mer}, J.},
  \bibinfo{author}{von Stechow, C.}, \bibinfo{author}{{Zwickel Senior
  Scientist}, T.}, \bibinfo{year}{2014}.
\newblock \bibinfo{title}{{Climate Change 2014 Mitigation of Climate Change
  Working Group III Contribution to the Fifth Assessment Report of the
  Intergovernmental Panel on Climate Change Edited by}}.
\newblock \bibinfo{type}{Technical Report}.
\newblock \URLprefix \url{www.cambridge.org}.
\bibitem[{Eising et~al.(2020)Eising, Hobbie and M{\"{o}}st}]{Eising2020}
\bibinfo{author}{Eising, M.}, \bibinfo{author}{Hobbie, H.},
  \bibinfo{author}{M{\"{o}}st, D.}, \bibinfo{year}{2020}.
\newblock \bibinfo{title}{{Future wind and solar power market values in Germany
  — Evidence of spatial and technological dependencies?}}
\newblock \bibinfo{journal}{Energy Econ.} \bibinfo{volume}{86},
  \bibinfo{pages}{104638}.
\newblock \DOIprefix\doi{10.1016/j.eneco.2019.104638}.
\bibitem[{Engeland et~al.(2017)Engeland, Borga, Creutin, Fran\c{c}ois, Ramos
  and Vidal}]{Engeland_2017}
\bibinfo{author}{Engeland, K.}, \bibinfo{author}{Borga, M.},
  \bibinfo{author}{Creutin, J.D.}, \bibinfo{author}{Fran\c{c}ois, B.},
  \bibinfo{author}{Ramos, M.H.}, \bibinfo{author}{Vidal, J.P.},
  \bibinfo{year}{2017}.
\newblock \bibinfo{title}{Space-time variability of climate variables and
  intermittent renewable electricity production – a review}.
\newblock \bibinfo{journal}{Renewable and Sustainable Energy Reviews}
  \bibinfo{volume}{79}, \bibinfo{pages}{600 -- 617}.
\newblock \DOIprefix\doi{10.1016/j.rser.2017.05.046}.
\bibitem[{Friedlingstein et~al.(2019)Friedlingstein, Jones, O'Sullivan, Andrew,
  Hauck, Peters, Peters, Pongratz, Sitch, {Le Qu{\'{e}}r{\'{e}}}, DBakker,
  Canadell1, Ciais1, Jackson, Anthoni1, Barbero, Bastos, Bastrikov, Becker,
  Bopp, Buitenhuis, Chandra, Chevallier, Chini, Currie, Feely, Gehlen,
  Gilfillan, Gkritzalis, Goll, Gruber, Gutekunst, Harris, Haverd, Houghton,
  Hurtt, Ilyina, Jain, Joetzjer, Kaplan, Kato, Goldewijk, Korsbakken,
  Landsch{\"{u}}tzer, Lauvset, Lef{\`{e}}vre, Lenton, Lienert, Lombardozzi,
  Marland, McGuire, Melton, Metzl, Munro, Nabel, Nakaoka, Neill, Omar, Ono,
  Peregon, Pierrot, Poulter, Rehder, Resplandy, Robertson, R{\"{o}}denbeck,
  S{\'{e}}f{\'{e}}rian, Schwinger, Smith, Tans, Tian, Tilbrook, Tubiello, {Van
  Der Werf}, Wiltshire and Zaehle}]{Friedlingstein2019}
\bibinfo{author}{Friedlingstein, P.}, \bibinfo{author}{Jones, M.W.},
  \bibinfo{author}{O'Sullivan, M.}, \bibinfo{author}{Andrew, R.M.},
  \bibinfo{author}{Hauck, J.}, \bibinfo{author}{Peters, G.P.},
  \bibinfo{author}{Peters, W.}, \bibinfo{author}{Pongratz, J.},
  \bibinfo{author}{Sitch, S.}, \bibinfo{author}{{Le Qu{\'{e}}r{\'{e}}}, C.},
  \bibinfo{author}{DBakker, O.C.}, \bibinfo{author}{Canadell1, J.G.},
  \bibinfo{author}{Ciais1, P.}, \bibinfo{author}{Jackson, R.B.},
  \bibinfo{author}{Anthoni1, P.}, \bibinfo{author}{Barbero, L.},
  \bibinfo{author}{Bastos, A.}, \bibinfo{author}{Bastrikov, V.},
  \bibinfo{author}{Becker, M.}, \bibinfo{author}{Bopp, L.},
  \bibinfo{author}{Buitenhuis, E.}, \bibinfo{author}{Chandra, N.},
  \bibinfo{author}{Chevallier, F.}, \bibinfo{author}{Chini, L.P.},
  \bibinfo{author}{Currie, K.I.}, \bibinfo{author}{Feely, R.A.},
  \bibinfo{author}{Gehlen, M.}, \bibinfo{author}{Gilfillan, D.},
  \bibinfo{author}{Gkritzalis, T.}, \bibinfo{author}{Goll, D.S.},
  \bibinfo{author}{Gruber, N.}, \bibinfo{author}{Gutekunst, S.},
  \bibinfo{author}{Harris, I.}, \bibinfo{author}{Haverd, V.},
  \bibinfo{author}{Houghton, R.A.}, \bibinfo{author}{Hurtt, G.},
  \bibinfo{author}{Ilyina, T.}, \bibinfo{author}{Jain, A.K.},
  \bibinfo{author}{Joetzjer, E.}, \bibinfo{author}{Kaplan, J.O.},
  \bibinfo{author}{Kato, E.}, \bibinfo{author}{Goldewijk, K.K.},
  \bibinfo{author}{Korsbakken, J.I.}, \bibinfo{author}{Landsch{\"{u}}tzer, P.},
  \bibinfo{author}{Lauvset, S.K.}, \bibinfo{author}{Lef{\`{e}}vre, N.},
  \bibinfo{author}{Lenton, A.}, \bibinfo{author}{Lienert, S.},
  \bibinfo{author}{Lombardozzi, D.}, \bibinfo{author}{Marland, G.},
  \bibinfo{author}{McGuire, P.C.}, \bibinfo{author}{Melton, J.R.},
  \bibinfo{author}{Metzl, N.}, \bibinfo{author}{Munro, D.R.},
  \bibinfo{author}{Nabel, J.E.}, \bibinfo{author}{Nakaoka, S.I.},
  \bibinfo{author}{Neill, C.}, \bibinfo{author}{Omar, A.M.},
  \bibinfo{author}{Ono, T.}, \bibinfo{author}{Peregon, A.},
  \bibinfo{author}{Pierrot, D.}, \bibinfo{author}{Poulter, B.},
  \bibinfo{author}{Rehder, G.}, \bibinfo{author}{Resplandy, L.},
  \bibinfo{author}{Robertson, E.}, \bibinfo{author}{R{\"{o}}denbeck, C.},
  \bibinfo{author}{S{\'{e}}f{\'{e}}rian, R.}, \bibinfo{author}{Schwinger, J.},
  \bibinfo{author}{Smith, N.}, \bibinfo{author}{Tans, P.P.},
  \bibinfo{author}{Tian, H.}, \bibinfo{author}{Tilbrook, B.},
  \bibinfo{author}{Tubiello, F.N.}, \bibinfo{author}{{Van Der Werf}, G.R.},
  \bibinfo{author}{Wiltshire, A.J.}, \bibinfo{author}{Zaehle, S.},
  \bibinfo{year}{2019}.
\newblock \bibinfo{title}{{Global carbon budget 2019}}.
\newblock \bibinfo{journal}{Earth Syst. Sci. Data} \bibinfo{volume}{11},
  \bibinfo{pages}{1783--1838}.
\newblock \DOIprefix\doi{10.5194/essd-11-1783-2019}.
\bibitem[{F\"{u}rsch et~al.(2013)F\"{u}rsch, Hagspiel, J\"{a}gemann, Nagl,
  Lindenberger and Tr\"{o}ster}]{Fuersch_2013}
\bibinfo{author}{F\"{u}rsch, M.}, \bibinfo{author}{Hagspiel, S.},
  \bibinfo{author}{J\"{a}gemann, C.}, \bibinfo{author}{Nagl, S.},
  \bibinfo{author}{Lindenberger, D.}, \bibinfo{author}{Tr\"{o}ster, E.},
  \bibinfo{year}{2013}.
\newblock \bibinfo{title}{The role of grid extensions in a cost-efficient
  transformation of the {E}uropean electricity system until 2050}.
\newblock \bibinfo{journal}{Applied Energy} \bibinfo{volume}{104},
  \bibinfo{pages}{642 -- 652}.
\newblock \DOIprefix\doi{10.1016/j.apenergy.2012.11.050}.
\bibitem[{Gaete-Morales et~al.(2020)Gaete-Morales, Kittel, Roth, Schill and
  Zerrahn}]{Gaete_2020}
\bibinfo{author}{Gaete-Morales, C.}, \bibinfo{author}{Kittel, M.},
  \bibinfo{author}{Roth, A.}, \bibinfo{author}{Schill, W.P.},
  \bibinfo{author}{Zerrahn, A.}, \bibinfo{year}{2020}.
\newblock \bibinfo{title}{{DIETERpy: a Python framework for The Dispatch and
  Investment Evaluation Tool with Endogenous Renewables}}.
\newblock \href{http://arxiv.org/abs/2010.00883}{{\tt arXiv:2010.00883}}.
\bibitem[{Gelabert et~al.(2011)Gelabert, Labandeira and Linares}]{Gelabert2011}
\bibinfo{author}{Gelabert, L.}, \bibinfo{author}{Labandeira, X.},
  \bibinfo{author}{Linares, P.}, \bibinfo{year}{2011}.
\newblock \bibinfo{title}{{An ex-post analysis of the effect of renewables and
  cogeneration on Spanish electricity prices}}.
\newblock \bibinfo{journal}{Energy Econ.} \bibinfo{volume}{33},
  \bibinfo{pages}{S59--S65}.
\newblock \DOIprefix\doi{10.1016/j.eneco.2011.07.027}.
\bibitem[{{German Advisory Council on Global Change}(2011)}]{wbgu2011}
\bibinfo{author}{{German Advisory Council on Global Change}},
  \bibinfo{year}{2011}.
\newblock \bibinfo{title}{{World in Transition. A Social Contract for
  Sustainability}}.
\newblock \bibinfo{type}{Technical Report}.
\newblock \URLprefix
  \url{https://www.wbgu.de/en/publications/publication/world-in-transition-a-social-contract-for-sustainability#section-downloads}.
\bibitem[{Green and Vasilakos(2011)}]{Green2011}
\bibinfo{author}{Green, R.}, \bibinfo{author}{Vasilakos, N.},
  \bibinfo{year}{2011}.
\newblock \bibinfo{title}{{The long-term impact of wind power on electricity
  prices and generating capacity}}, in: \bibinfo{booktitle}{IEEE Power Energy
  Soc. Gen. Meet.}
\newblock \DOIprefix\doi{10.1109/PES.2011.6039218}.
\bibitem[{Green and L{\'{e}}autier(2015)}]{Green2015}
\bibinfo{author}{Green, R.J.}, \bibinfo{author}{L{\'{e}}autier, T.O.},
  \bibinfo{year}{2015}.
\newblock \bibinfo{title}{{Do costs fall faster than revenues ? Dynamics of
  renewables entry into electricity markets}}.
\newblock \bibinfo{type}{TSE Working Papers} \bibinfo{number}{15-591}. Toulouse
  School of Economics.
\newblock \URLprefix \url{https://ideas.repec.org/p/tse/wpaper/29543.html}.
\bibitem[{Grossmann et~al.(2014)Grossmann, Grossmann and
  Steininger}]{Grossmann2014}
\bibinfo{author}{Grossmann, W.D.}, \bibinfo{author}{Grossmann, I.},
  \bibinfo{author}{Steininger, K.W.}, \bibinfo{year}{2014}.
\newblock \bibinfo{title}{{Solar electricity generation across large geographic
  areas, Part II: A Pan-American energy system based on solar}}.
\newblock \bibinfo{journal}{Renew. Sustain. Energy Rev.} \bibinfo{volume}{32},
  \bibinfo{pages}{983--993}.
\newblock \DOIprefix\doi{10.1016/j.rser.2014.01.003}.
\bibitem[{G\"{u}r(2018)}]{Guer_2018}
\bibinfo{author}{G\"{u}r, T.M.}, \bibinfo{year}{2018}.
\newblock \bibinfo{title}{Review of electrical energy storage technologies{,}
  materials and systems: challenges and prospects for large-scale grid
  storage}.
\newblock \bibinfo{journal}{Energy \& Environmental Science}
  \bibinfo{volume}{11}, \bibinfo{pages}{2696--2767}.
\newblock \DOIprefix\doi{10.1039/C8EE01419A}.
\bibitem[{Hale et~al.(2018)Hale, Bird, Padmanabhan and Volpi}]{Hale2018}
\bibinfo{author}{Hale, E.}, \bibinfo{author}{Bird, L.},
  \bibinfo{author}{Padmanabhan, R.}, \bibinfo{author}{Volpi, C.},
  \bibinfo{year}{2018}.
\newblock \bibinfo{title}{{Potential Roles for Demand Response in High-Growth
  Electric Systems with Increasing Shares of Renewable Generation}}.
\newblock \bibinfo{type}{Technical Report}.
\newblock \URLprefix \url{www.nrel.gov/publications.}
\bibitem[{Heide et~al.(2010)Heide, {von Bremen}, Greiner, Hoffmann, Speckmann
  and Bofinger}]{Heide_2010}
\bibinfo{author}{Heide, D.}, \bibinfo{author}{{von Bremen}, L.},
  \bibinfo{author}{Greiner, M.}, \bibinfo{author}{Hoffmann, C.},
  \bibinfo{author}{Speckmann, M.}, \bibinfo{author}{Bofinger, S.},
  \bibinfo{year}{2010}.
\newblock \bibinfo{title}{Seasonal optimal mix of wind and solar power in a
  future, highly renewable europe}.
\newblock \bibinfo{journal}{Renewable Energy} \bibinfo{volume}{35},
  \bibinfo{pages}{2483 -- 2489}.
\newblock \DOIprefix\doi{10.1016/j.renene.2010.03.012}.
\bibitem[{Heuberger et~al.(2017)Heuberger, Staffell, Shah and
  Dowell}]{Heuberger_2017}
\bibinfo{author}{Heuberger, C.F.}, \bibinfo{author}{Staffell, I.},
  \bibinfo{author}{Shah, N.}, \bibinfo{author}{Dowell, N.M.},
  \bibinfo{year}{2017}.
\newblock \bibinfo{title}{A systems approach to quantifying the value of power
  generation and energy storage technologies in future electricity networks}.
\newblock \bibinfo{journal}{Computers \& Chemical Engineering}
  \bibinfo{volume}{107}, \bibinfo{pages}{247 -- 256}.
\newblock \DOIprefix\doi{10.1016/j.compchemeng.2017.05.012}.
\bibitem[{Hirth(2013)}]{Hirth2013a}
\bibinfo{author}{Hirth, L.}, \bibinfo{year}{2013}.
\newblock \bibinfo{title}{{The market value of variable renewables. The effect
  of solar wind power variability on their relative price}}.
\newblock \bibinfo{journal}{Energy Econ.} \bibinfo{volume}{38},
  \bibinfo{pages}{218--236}.
\newblock \DOIprefix\doi{10.1016/j.eneco.2013.02.004}.
\bibitem[{Hirth(2015)}]{Hirth2015d}
\bibinfo{author}{Hirth, L.}, \bibinfo{year}{2015}.
\newblock \bibinfo{title}{{The optimal share of variable renewables: How the
  variability of wind and solar power affects their welfare-optimal
  deployment}}.
\newblock \bibinfo{journal}{Energy J.} \bibinfo{volume}{36},
  \bibinfo{pages}{149--184}.
\newblock \DOIprefix\doi{10.5547/01956574.36.1.6}.
\bibitem[{Hirth et~al.(2015)Hirth, Ueckerdt and Edenhofer}]{Hirth2015e}
\bibinfo{author}{Hirth, L.}, \bibinfo{author}{Ueckerdt, F.},
  \bibinfo{author}{Edenhofer, O.}, \bibinfo{year}{2015}.
\newblock \bibinfo{title}{{Integration costs revisited - An economic framework
  for wind and solar variability}}.
\newblock \bibinfo{journal}{Renew. Energy} \bibinfo{volume}{74},
  \bibinfo{pages}{925--939}.
\newblock \DOIprefix\doi{10.1016/j.renene.2014.08.065}.
\bibitem[{Hirth et~al.(2016)Hirth, Ueckerdt and Edenhofer}]{Hirth2016}
\bibinfo{author}{Hirth, L.}, \bibinfo{author}{Ueckerdt, F.},
  \bibinfo{author}{Edenhofer, O.}, \bibinfo{year}{2016}.
\newblock \bibinfo{title}{{Why wind is not coal: On the economics of
  electricity generation}}.
\newblock \bibinfo{journal}{Energy J.} \bibinfo{volume}{37},
  \bibinfo{pages}{1--27}.
\newblock \DOIprefix\doi{10.5547/01956574.37.3.lhir}.
\bibitem[{Hoekstra et~al.(2017)Hoekstra, Steinbuch and Verbong}]{Hoekstra2017}
\bibinfo{author}{Hoekstra, A.}, \bibinfo{author}{Steinbuch, M.},
  \bibinfo{author}{Verbong, G.}, \bibinfo{year}{2017}.
\newblock \bibinfo{title}{{Creating agent-based energy transition management
  models that can uncover profitable pathways to climate change mitigation}}.
\newblock \bibinfo{journal}{Complexity} \bibinfo{volume}{2017}.
\newblock \DOIprefix\doi{10.1155/2017/1967645}.
\bibitem[{{IEA}(2017)}]{IEA2020Digital}
\bibinfo{author}{{IEA}}, \bibinfo{year}{2017}.
\newblock \bibinfo{title}{{Digitalization {\&} Energy}}.
\newblock \bibinfo{type}{Technical Report}. International Energy Agency.
\newblock \URLprefix \url{www.iea.org/t{\&}c/}.
\bibitem[{{IEA}(2018)}]{IEA2018WEO}
\bibinfo{author}{{IEA}}, \bibinfo{year}{2018}.
\newblock \bibinfo{title}{{World Energy Outlook 2018}}.
\newblock \bibinfo{type}{Technical Report}. International Energy Agency.
\newblock \URLprefix
  \url{https://www.iea.org/reports/world-energy-outlook-2018}.
\bibitem[{{IEA}(2019a)}]{IEA2019H2}
\bibinfo{author}{{IEA}}, \bibinfo{year}{2019}a.
\newblock \bibinfo{title}{{The Future of Hydrogen}}.
\newblock \bibinfo{type}{Technical Report}. International Energy Agency.
\newblock \URLprefix \url{https://www.iea.org/reports/the-future-of-hydrogen},
  \DOIprefix\doi{10.1787/1e0514c4-en}.
\bibitem[{{IEA}(2019b)}]{IEA2019WEO}
\bibinfo{author}{{IEA}}, \bibinfo{year}{2019}b.
\newblock \bibinfo{title}{{World Energy Outlook 2019}}.
\newblock \bibinfo{type}{Technical Report}. International Energy Agency.
\newblock \URLprefix
  \url{https://www.iea.org/reports/world-energy-outlook-2019}.
  \bibinfo{note}{international Energy Agency}.
\bibitem[{{IEA}(2020a)}]{IEA2020ETP}
\bibinfo{author}{{IEA}}, \bibinfo{year}{2020}a.
\newblock \bibinfo{title}{{Energy Technology Perspectives 2020}}.
\newblock \bibinfo{type}{Technical Report}. International Energy Agency.
\newblock \URLprefix \url{www.iea.org/t{\&}c/}.
\bibitem[{{IEA}(2020b)}]{IEA2020Investment}
\bibinfo{author}{{IEA}}, \bibinfo{year}{2020}b.
\newblock \bibinfo{title}{{World Energy Investment 2020}}.
\newblock \bibinfo{type}{Technical Report}. International Energy Agency.
\newblock \URLprefix \url{www.iea.org/t{\&}c/}.
\bibitem[{IRENA(2020)}]{IRENA2020}
\bibinfo{author}{IRENA}, \bibinfo{year}{2020}.
\newblock \bibinfo{title}{{Renewable Power Generation Costs in 2019}}.
\newblock \bibinfo{type}{Technical Report}. International Renewable Energy
  Agency. \bibinfo{address}{Abu Dhabi}.
\newblock \URLprefix
  \url{https://www.irena.org/-/media/Files/IRENA/Agency/Publication/2018/Jan/IRENA_2017_Power_Costs_2018.pdf}.
\bibitem[{Jacobson et~al.(2017)Jacobson, Delucchi, Bauer, Goodman, Chapman,
  Cameron, Bozonnat, Chobadi, Clonts, Enevoldsen, Erwin, Fobi, Goldstrom,
  Hennessy, Liu, Lo, Meyer, Morris, Moy, O'Neill, Petkov, Redfern, Schucker,
  Sontag, Wang, Weiner and Yachanin}]{Jacobson_2017}
\bibinfo{author}{Jacobson, M.Z.}, \bibinfo{author}{Delucchi, M.A.},
  \bibinfo{author}{Bauer, Z.A.}, \bibinfo{author}{Goodman, S.C.},
  \bibinfo{author}{Chapman, W.E.}, \bibinfo{author}{Cameron, M.A.},
  \bibinfo{author}{Bozonnat, C.}, \bibinfo{author}{Chobadi, L.},
  \bibinfo{author}{Clonts, H.A.}, \bibinfo{author}{Enevoldsen, P.},
  \bibinfo{author}{Erwin, J.R.}, \bibinfo{author}{Fobi, S.N.},
  \bibinfo{author}{Goldstrom, O.K.}, \bibinfo{author}{Hennessy, E.M.},
  \bibinfo{author}{Liu, J.}, \bibinfo{author}{Lo, J.}, \bibinfo{author}{Meyer,
  C.B.}, \bibinfo{author}{Morris, S.B.}, \bibinfo{author}{Moy, K.R.},
  \bibinfo{author}{O'Neill, P.L.}, \bibinfo{author}{Petkov, I.},
  \bibinfo{author}{Redfern, S.}, \bibinfo{author}{Schucker, R.},
  \bibinfo{author}{Sontag, M.A.}, \bibinfo{author}{Wang, J.},
  \bibinfo{author}{Weiner, E.}, \bibinfo{author}{Yachanin, A.S.},
  \bibinfo{year}{2017}.
\newblock \bibinfo{title}{100\% clean and renewable wind, water, and sunlight
  all-sector energy roadmaps for 139 countries of the world}.
\newblock \bibinfo{journal}{Joule} \bibinfo{volume}{1}, \bibinfo{pages}{108 --
  121}.
\newblock \DOIprefix\doi{10.1016/j.joule.2017.07.005}.
\bibitem[{Joskow(2011)}]{Joskow2011a}
\bibinfo{author}{Joskow, B.P.L.}, \bibinfo{year}{2011}.
\newblock \bibinfo{title}{{Comparing the Costs of Intermittent and Dispatchable
  Electricity Generating Technologies}}.
\newblock \bibinfo{journal}{Am. Econ. Rev. Pap. Proc.} \bibinfo{volume}{101}.
\bibitem[{Joskow(2019)}]{Joskow2019}
\bibinfo{author}{Joskow, P.L.}, \bibinfo{year}{2019}.
\newblock \bibinfo{title}{{Challenges for wholesale electricity markets with
  intermittent renewable generation at scale: The US experience}}.
\newblock \bibinfo{journal}{Oxford Rev. Econ. Policy} \bibinfo{volume}{35},
  \bibinfo{pages}{291--331}.
\newblock \DOIprefix\doi{10.1093/oxrep/grz001}.
\bibitem[{Karkour et~al.(2020)Karkour, Ichisugi, Abeynayaka and
  Itsubo}]{Karkour2020}
\bibinfo{author}{Karkour, S.}, \bibinfo{author}{Ichisugi, Y.},
  \bibinfo{author}{Abeynayaka, A.}, \bibinfo{author}{Itsubo, N.},
  \bibinfo{year}{2020}.
\newblock \bibinfo{title}{{External-cost estimation of electricity generation
  in G20 countries: Case study using a global life-cycle impact-assessment
  method}}.
\newblock \bibinfo{journal}{Sustain.} \bibinfo{volume}{12}.
\newblock \DOIprefix\doi{10.3390/su12052002}.
\bibitem[{Kempton and Tomi\'{c}(2005)}]{Kempton_2005}
\bibinfo{author}{Kempton, W.}, \bibinfo{author}{Tomi\'{c}, J.},
  \bibinfo{year}{2005}.
\newblock \bibinfo{title}{Vehicle-to-grid power fundamentals: Calculating
  capacity and net revenue}.
\newblock \bibinfo{journal}{Journal of Power Sources} \bibinfo{volume}{144},
  \bibinfo{pages}{268 -- 279}.
\newblock \DOIprefix\doi{10.1016/j.jpowsour.2004.12.025}.
\bibitem[{van Kooten(2016)}]{VanKooten2016}
\bibinfo{author}{van Kooten, G.C.}, \bibinfo{year}{2016}.
\newblock \bibinfo{title}{{The Economics of Wind Power}}.
\newblock \bibinfo{journal}{Annu. Rev. Resour. Econ.} \bibinfo{volume}{8},
  \bibinfo{pages}{181--205}.
\newblock \DOIprefix\doi{10.1146/annurev-resource-091115-022544}.
\bibitem[{Kyritsis et~al.(2017)Kyritsis, Andersson and Serletis}]{Kyritsis2017}
\bibinfo{author}{Kyritsis, E.}, \bibinfo{author}{Andersson, J.},
  \bibinfo{author}{Serletis, A.}, \bibinfo{year}{2017}.
\newblock \bibinfo{title}{{Electricity prices, large-scale renewable
  integration, and policy implications}}.
\newblock \bibinfo{journal}{Energy Policy} \bibinfo{volume}{101},
  \bibinfo{pages}{550--560}.
\newblock \DOIprefix\doi{10.1016/j.enpol.2016.11.014}.
\bibitem[{Lamont(2008)}]{Lamont2008}
\bibinfo{author}{Lamont, A.D.}, \bibinfo{year}{2008}.
\newblock \bibinfo{title}{{Assessing the long-term system value of intermittent
  electric generation technologies}}.
\newblock \bibinfo{journal}{Energy Econ.} \bibinfo{volume}{30},
  \bibinfo{pages}{1208--1231}.
\newblock \DOIprefix\doi{10.1016/j.eneco.2007.02.007}.
\bibitem[{Lang et~al.(2016)Lang, Ammann and Girod}]{Lang2016a}
\bibinfo{author}{Lang, T.}, \bibinfo{author}{Ammann, D.},
  \bibinfo{author}{Girod, B.}, \bibinfo{year}{2016}.
\newblock \bibinfo{title}{{Profitability in absence of subsidies: A
  techno-economic analysis of rooftop photovoltaic self-consumption in
  residential and commercial buildings}}.
\newblock \bibinfo{journal}{Renewable Energy} \bibinfo{volume}{87},
  \bibinfo{pages}{77--87}.
\newblock \DOIprefix\doi{10.1016/j.renene.2015.09.059}.
\bibitem[{Lazard(2019)}]{Lazard2019a}
\bibinfo{author}{Lazard}, \bibinfo{year}{2019}.
\newblock \bibinfo{title}{{Lazard's levelized cost of storage analysis version
  5.0}}.
\newblock \bibinfo{type}{Technical Report}.
\newblock \URLprefix \url{https://www.lazard.com/perspective/lcoe2019}.
\bibitem[{L{\'{e}}autier(2019)}]{Leautier2019}
\bibinfo{author}{L{\'{e}}autier, T.O.}, \bibinfo{year}{2019}.
\newblock \bibinfo{title}{{Imperfect Markets and Imperfect Regulation: An
  Introduction to the Microeconomics and Political Economy of Power Markets}}.
\newblock \URLprefix
  \url{https://mitpress.mit.edu/books/imperfect-markets-and-imperfect-regulation}.
\bibitem[{Leslie et~al.(2020)Leslie, Stern, Shanker and Hogan}]{Leslie2020}
\bibinfo{author}{Leslie, G.W.}, \bibinfo{author}{Stern, D.I.},
  \bibinfo{author}{Shanker, A.}, \bibinfo{author}{Hogan, M.T.},
  \bibinfo{year}{2020}.
\newblock \bibinfo{title}{{Designing electricity markets for high penetrations
  of zero or low marginal cost intermittent energy sources}}.
\newblock \bibinfo{journal}{Electr. J.} \bibinfo{volume}{33},
  \bibinfo{pages}{106847}.
\newblock \DOIprefix\doi{10.1016/j.tej.2020.106847}.
\bibitem[{{L{\'{o}}pez Prol} and Steininger(2020)}]{LopezProl2020a}
\bibinfo{author}{{L{\'{o}}pez Prol}, J.}, \bibinfo{author}{Steininger, K.W.},
  \bibinfo{year}{2020}.
\newblock \bibinfo{title}{{Photovoltaic self-consumption is now profitable in
  Spain. Effects of the new regulation on prosumers' internal rate of return}}.
\newblock \bibinfo{journal}{Energy Policy} \bibinfo{volume}{146},
  \bibinfo{pages}{111793}.
\newblock \DOIprefix\doi{10.1016/j.enpol.2020.111793}.
\bibitem[{{L{\'{o}}pez Prol} et~al.(2020){L{\'{o}}pez Prol}, Steininger and
  Zilberman}]{LopezProl2020}
\bibinfo{author}{{L{\'{o}}pez Prol}, J.}, \bibinfo{author}{Steininger, K.W.},
  \bibinfo{author}{Zilberman, D.}, \bibinfo{year}{2020}.
\newblock \bibinfo{title}{{The cannibalization effect of wind and solar in the
  California wholesale electricity market}}.
\newblock \bibinfo{journal}{Energy Econ.} \bibinfo{volume}{85},
  \bibinfo{pages}{104552}.
\newblock \DOIprefix\doi{10.1016/j.eneco.2019.104552}.
\bibitem[{Lund et~al.(2015)Lund, Lindgren, Mikkola and Salpakari}]{Lund_2015}
\bibinfo{author}{Lund, P.D.}, \bibinfo{author}{Lindgren, J.},
  \bibinfo{author}{Mikkola, J.}, \bibinfo{author}{Salpakari, J.},
  \bibinfo{year}{2015}.
\newblock \bibinfo{title}{Review of energy system flexibility measures to
  enable high levels of variable renewable electricity}.
\newblock \bibinfo{journal}{Renewable and Sustainable Energy Reviews}
  \bibinfo{volume}{45}, \bibinfo{pages}{785 -- 807}.
\newblock \DOIprefix\doi{10.1016/j.rser.2015.01.057}.
\bibitem[{Luo et~al.(2015)Luo, Wang, Dooner and Clarke}]{Luo_2015}
\bibinfo{author}{Luo, X.}, \bibinfo{author}{Wang, J.}, \bibinfo{author}{Dooner,
  M.}, \bibinfo{author}{Clarke, J.}, \bibinfo{year}{2015}.
\newblock \bibinfo{title}{Overview of current development in electrical energy
  storage technologies and the application potential in power system
  operation}.
\newblock \bibinfo{journal}{Applied Energy} \bibinfo{volume}{137},
  \bibinfo{pages}{511 -- 536}.
\newblock \DOIprefix\doi{10.1016/j.apenergy.2014.09.081}.
\bibitem[{MacDonald et~al.(2016)MacDonald, Clack, Alexander, Dunbar, Wilczak
  and Xie}]{MacDonald_2016}
\bibinfo{author}{MacDonald, A.E.}, \bibinfo{author}{Clack, C.T.M.},
  \bibinfo{author}{Alexander, A.}, \bibinfo{author}{Dunbar, A.},
  \bibinfo{author}{Wilczak, J.}, \bibinfo{author}{Xie, Y.},
  \bibinfo{year}{2016}.
\newblock \bibinfo{title}{Future cost-competitive electricity systems and their
  impact on {US} {CO}$_{2}$ emissions}.
\newblock \bibinfo{journal}{Nature Climate Change} \bibinfo{volume}{6},
  \bibinfo{pages}{526--531}.
\newblock \DOIprefix\doi{10.1038/nclimate2921}.
\bibitem[{Madeddu et~al.(2020)Madeddu, Ueckerdt, Pehl, Peterseim, Lord, Kumar,
  Kr\"{u}ger and Luderer}]{Madeddu_2020}
\bibinfo{author}{Madeddu, S.}, \bibinfo{author}{Ueckerdt, F.},
  \bibinfo{author}{Pehl, M.}, \bibinfo{author}{Peterseim, J.},
  \bibinfo{author}{Lord, M.}, \bibinfo{author}{Kumar, K.A.},
  \bibinfo{author}{Kr\"{u}ger, C.}, \bibinfo{author}{Luderer, G.},
  \bibinfo{year}{2020}.
\newblock \bibinfo{title}{The co$_2$ reduction potential for the {European}
  industry via direct electrification of heat supply (power-to-heat)}.
\newblock \bibinfo{journal}{Environmental Research Letters}
  \bibinfo{volume}{15}, \bibinfo{pages}{124004}.
\newblock \DOIprefix\doi{10.1088/1748-9326/abbd02}.
\bibitem[{Mallapragada et~al.(2020)Mallapragada, Sepulveda and
  Jenkins}]{Mallapragada_2020}
\bibinfo{author}{Mallapragada, D.S.}, \bibinfo{author}{Sepulveda, N.A.},
  \bibinfo{author}{Jenkins, J.D.}, \bibinfo{year}{2020}.
\newblock \bibinfo{title}{Long-run system value of battery energy storage in
  future grids with increasing wind and solar generation}.
\newblock \bibinfo{journal}{Applied Energy} \bibinfo{volume}{275},
  \bibinfo{pages}{115390}.
\newblock \DOIprefix\doi{10.1016/j.apenergy.2020.115390}.
\bibitem[{Mansilla et~al.(2018)Mansilla, Bourasseau, Cany, Guinot, Duigou and
  Lucchese}]{Mansilla2018}
\bibinfo{author}{Mansilla, C.}, \bibinfo{author}{Bourasseau, C.},
  \bibinfo{author}{Cany, C.}, \bibinfo{author}{Guinot, B.},
  \bibinfo{author}{Duigou, A.L.}, \bibinfo{author}{Lucchese, P.},
  \bibinfo{year}{2018}.
\newblock \bibinfo{title}{{Hydrogen applications: Overview of the key economic
  issues and perspectives}}, in: \bibinfo{booktitle}{Hydrog. Supply Chain Des.
  Deploy. Oper.}. \bibinfo{publisher}{Elsevier}, pp. \bibinfo{pages}{271--292}.
\newblock \DOIprefix\doi{10.1016/B978-0-12-811197-0.00007-5}.
\bibitem[{Masson-Delmotte et~al.(2018)Masson-Delmotte, Zhai, P{\"{o}}rtner,
  Roberts, Skea, Shukla, Pirani, Moufouma-Okia, P{\'{e}}an, Pidcock, Connors,
  Matthews, Chen, Zhou, Gomis, Lonnoy, Maycock, Tignor and
  Waterfield}]{IPCC2018SR1.5}
\bibinfo{author}{Masson-Delmotte, V.}, \bibinfo{author}{Zhai, P.},
  \bibinfo{author}{P{\"{o}}rtner, H.O.}, \bibinfo{author}{Roberts, D.},
  \bibinfo{author}{Skea, J.}, \bibinfo{author}{Shukla, P.R.},
  \bibinfo{author}{Pirani, A.}, \bibinfo{author}{Moufouma-Okia, W.},
  \bibinfo{author}{P{\'{e}}an, C.}, \bibinfo{author}{Pidcock, R.},
  \bibinfo{author}{Connors, S.}, \bibinfo{author}{Matthews, J.B.R.},
  \bibinfo{author}{Chen, Y.}, \bibinfo{author}{Zhou, X.},
  \bibinfo{author}{Gomis, M.I.}, \bibinfo{author}{Lonnoy, E.},
  \bibinfo{author}{Maycock, T.}, \bibinfo{author}{Tignor, M.},
  \bibinfo{author}{Waterfield, T.}, \bibinfo{year}{2018}.
\newblock \bibinfo{title}{{Global warming of 1.5$^{\circ}${C}. An IPCC Special
  Report on the impacts of global warming of 1.5$^{\circ}${C} above
  pre-industrial levels and related global greenhouse gas emission pathways, in
  the context of strengthening the global response to the threat of climate
  change}}.
\newblock \bibinfo{type}{Technical Report}.
\bibitem[{Mayer et~al.(2015)Mayer, Philipps, Hussein, Schlegl and
  Senkpiel}]{Mayer2015}
\bibinfo{author}{Mayer, J.N.}, \bibinfo{author}{Philipps, S.},
  \bibinfo{author}{Hussein, N.S.}, \bibinfo{author}{Schlegl, T.},
  \bibinfo{author}{Senkpiel, C.}, \bibinfo{year}{2015}.
\newblock \bibinfo{title}{{Current and Future Cost of Photovoltaics Long-term
  Scenarios for Market Development, System Prices and LCOE of Utility-Scale PV
  Systems}}.
\newblock \bibinfo{type}{Technical Report}. Agpora Energiewende.
\newblock \URLprefix
  \url{https://www.ise.fraunhofer.de/content/dam/ise/de/documents/publications/studies/AgoraEnergiewende_Current_and_Future_Cost_of_PV_Feb2015_web.pdf}.
\bibitem[{McConnell et~al.(2015)McConnell, Forcey and
  Sandiford}]{McConnell_2015}
\bibinfo{author}{McConnell, D.}, \bibinfo{author}{Forcey, T.},
  \bibinfo{author}{Sandiford, M.}, \bibinfo{year}{2015}.
\newblock \bibinfo{title}{Estimating the value of electricity storage in an
  energy-only wholesale market}.
\newblock \bibinfo{journal}{Applied Energy} \bibinfo{volume}{159},
  \bibinfo{pages}{422 -- 432}.
\newblock \DOIprefix\doi{10.1016/j.apenergy.2015.09.006}.
\bibitem[{Mills and Wiser(2014)}]{Mills2014}
\bibinfo{author}{Mills, A.}, \bibinfo{author}{Wiser, R.}, \bibinfo{year}{2014}.
\newblock \bibinfo{title}{{Strategies for Mitigating the Reduction in Economic
  Value of Variable Generation with Increasing Penetration Levels}}.
\newblock \bibinfo{journal}{Ernest Orlando Lawrence Berkeley Natl. Lab.}
  \bibinfo{volume}{36}, \bibinfo{pages}{119--127}.
\newblock \DOIprefix\doi{10.1016/j.exphem.2007.09.002}.
\bibitem[{Mills et~al.(2020)Mills, Levin, Wiser, Seel and Botterud}]{Mills2020}
\bibinfo{author}{Mills, A.D.}, \bibinfo{author}{Levin, T.},
  \bibinfo{author}{Wiser, R.}, \bibinfo{author}{Seel, J.},
  \bibinfo{author}{Botterud, A.}, \bibinfo{year}{2020}.
\newblock \bibinfo{title}{{Impacts of variable renewable energy on wholesale
  markets and generating assets in the United States: A review of expectations
  and evidence}}.
\newblock \bibinfo{journal}{Renew. Sustain. Energy Rev.} \bibinfo{volume}{120},
  \bibinfo{pages}{109670}.
\newblock \DOIprefix\doi{10.1016/j.rser.2019.109670}.
\bibitem[{Mills and Wiser(2012)}]{Mills2012a}
\bibinfo{author}{Mills, A.D.}, \bibinfo{author}{Wiser, R.H.},
  \bibinfo{year}{2012}.
\newblock \bibinfo{title}{{Changes in the economic value of photovoltaic
  generation at high penetration levels: A pilot case study of California}},
  in: \bibinfo{booktitle}{Conf. Rec. IEEE Photovolt. Spec. Conf.}
\newblock \DOIprefix\doi{10.1109/PVSC-Vol2.2013.6656763}.
\bibitem[{Palizban and Kauhaniemi(2016)}]{Palizban_2016}
\bibinfo{author}{Palizban, O.}, \bibinfo{author}{Kauhaniemi, K.},
  \bibinfo{year}{2016}.
\newblock \bibinfo{title}{Energy storage systems in modern grids -- matrix of
  technologies and applications}.
\newblock \bibinfo{journal}{Journal of Energy Storage} \bibinfo{volume}{6},
  \bibinfo{pages}{248 -- 259}.
\newblock \DOIprefix\doi{10.1016/j.est.2016.02.001}.
\bibitem[{Rai and Nunn(2020)}]{Rai2020}
\bibinfo{author}{Rai, A.}, \bibinfo{author}{Nunn, O.}, \bibinfo{year}{2020}.
\newblock \bibinfo{title}{{On the impact of increasing penetration of variable
  renewables on electricity spot price extremes in Australia}}.
\newblock \bibinfo{journal}{Econ. Anal. Policy} \bibinfo{volume}{67},
  \bibinfo{pages}{67--86}.
\newblock \DOIprefix\doi{10.1016/j.eap.2020.06.001}.
\bibitem[{Rasmussen et~al.(2012)Rasmussen, Andresen and
  Greiner}]{Rasmussen_2012}
\bibinfo{author}{Rasmussen, M.G.}, \bibinfo{author}{Andresen, G.B.},
  \bibinfo{author}{Greiner, M.}, \bibinfo{year}{2012}.
\newblock \bibinfo{title}{Storage and balancing synergies in a fully or highly
  renewable pan-{E}uropean power system}.
\newblock \bibinfo{journal}{Energy Policy} \bibinfo{volume}{51},
  \bibinfo{pages}{642 -- 651}.
\newblock \DOIprefix\doi{10.1016/j.enpol.2012.09.009}.
\bibitem[{Raynaud et~al.(2018)Raynaud, Hingray, Fran\c{c}ois and
  Creutin}]{Raynaud_2018}
\bibinfo{author}{Raynaud, D.}, \bibinfo{author}{Hingray, B.},
  \bibinfo{author}{Fran\c{c}ois, B.}, \bibinfo{author}{Creutin, J.},
  \bibinfo{year}{2018}.
\newblock \bibinfo{title}{Energy droughts from variable renewable energy
  sources in {European} climates}.
\newblock \bibinfo{journal}{Renewable Energy} \bibinfo{volume}{125},
  \bibinfo{pages}{578 -- 589}.
\newblock \DOIprefix\doi{10.1016/j.renene.2018.02.130}.
\bibitem[{Rintam{\"{a}}ki et~al.(2017)Rintam{\"{a}}ki, Siddiqui and
  Salo}]{Rintamaki2017}
\bibinfo{author}{Rintam{\"{a}}ki, T.}, \bibinfo{author}{Siddiqui, A.S.},
  \bibinfo{author}{Salo, A.}, \bibinfo{year}{2017}.
\newblock \bibinfo{title}{{Does renewable energy generation decrease the
  volatility of electricity prices? An analysis of Denmark and Germany}}.
\newblock \bibinfo{journal}{Energy Econ.} \bibinfo{volume}{62},
  \bibinfo{pages}{270--282}.
\newblock \DOIprefix\doi{10.1016/j.eneco.2016.12.019}.
\bibitem[{Runge et~al.(2019)Runge, S\"{o}lch, Albert, Wasserscheid, Z\"{o}ttl
  and Grimm}]{Runge_2019}
\bibinfo{author}{Runge, P.}, \bibinfo{author}{S\"{o}lch, C.},
  \bibinfo{author}{Albert, J.}, \bibinfo{author}{Wasserscheid, P.},
  \bibinfo{author}{Z\"{o}ttl, G.}, \bibinfo{author}{Grimm, V.},
  \bibinfo{year}{2019}.
\newblock \bibinfo{title}{Economic comparison of different electric fuels for
  energy scenarios in 2035}.
\newblock \bibinfo{journal}{Applied Energy} \bibinfo{volume}{233-234},
  \bibinfo{pages}{1078--1093}.
\newblock \DOIprefix\doi{10.1016/j.apenergy.2018.10.023}.
\bibitem[{{S{\'{a}}enz de Miera} et~al.(2008){S{\'{a}}enz de Miera}, {del
  R{\'{i}}o Gonz{\'{a}}lez} and Vizca{\'{i}}no}]{SaenzdeMiera2008}
\bibinfo{author}{{S{\'{a}}enz de Miera}, G.}, \bibinfo{author}{{del R{\'{i}}o
  Gonz{\'{a}}lez}, P.}, \bibinfo{author}{Vizca{\'{i}}no, I.},
  \bibinfo{year}{2008}.
\newblock \bibinfo{title}{{Analysing the impact of renewable electricity
  support schemes on power prices: The case of wind electricity in Spain}}.
\newblock \bibinfo{journal}{Energy Policy} \bibinfo{volume}{36},
  \bibinfo{pages}{3345--3359}.
\newblock \DOIprefix\doi{10.1016/J.ENPOL.2008.04.022}.
\bibitem[{Safaei and Keith(2015)}]{Safaei_2015}
\bibinfo{author}{Safaei, H.}, \bibinfo{author}{Keith, D.W.},
  \bibinfo{year}{2015}.
\newblock \bibinfo{title}{How much bulk energy storage is needed to decarbonize
  electricity?}
\newblock \bibinfo{journal}{Energy \& Environmental Science}
  \bibinfo{volume}{8}, \bibinfo{pages}{3409--3417}.
\newblock \DOIprefix\doi{10.1039/C5EE01452B}.
\bibitem[{Say et~al.(2020)Say, Schill and John}]{Say_2020}
\bibinfo{author}{Say, K.}, \bibinfo{author}{Schill, W.P.},
  \bibinfo{author}{John, M.}, \bibinfo{year}{2020}.
\newblock \bibinfo{title}{Degrees of displacement: The impact of household {PV}
  battery prosumage on utility generation and storage}.
\newblock \bibinfo{journal}{Applied Energy} \bibinfo{volume}{276},
  \bibinfo{pages}{115466}.
\newblock \DOIprefix\doi{10.1016/j.apenergy.2020.115466}.
\bibitem[{Schiebahn et~al.(2015)Schiebahn, Grube, Robinius, Tietze, Kumar and
  Stolten}]{Schiebahn_2015}
\bibinfo{author}{Schiebahn, S.}, \bibinfo{author}{Grube, T.},
  \bibinfo{author}{Robinius, M.}, \bibinfo{author}{Tietze, V.},
  \bibinfo{author}{Kumar, B.}, \bibinfo{author}{Stolten, D.},
  \bibinfo{year}{2015}.
\newblock \bibinfo{title}{Power to gas: Technological overview, systems
  analysis and economic assessment for a case study in germany}.
\newblock \bibinfo{journal}{International Journal of Hydrogen Energy}
  \bibinfo{volume}{40}, \bibinfo{pages}{4285 -- 4294}.
\newblock \DOIprefix\doi{10.1016/j.ijhydene.2015.01.123}.
\bibitem[{Schill(2014)}]{Schill_2014}
\bibinfo{author}{Schill, W.P.}, \bibinfo{year}{2014}.
\newblock \bibinfo{title}{Residual load, renewable surplus generation and
  storage requirements in {Germany}}.
\newblock \bibinfo{journal}{Energy Policy} \bibinfo{volume}{73},
  \bibinfo{pages}{65 -- 79}.
\newblock \DOIprefix\doi{10.1016/j.enpol.2014.05.032}.
\bibitem[{Schill(2020)}]{Schill_Joule_2020}
\bibinfo{author}{Schill, W.P.}, \bibinfo{year}{2020}.
\newblock \bibinfo{title}{Electricity storage and the renewable energy
  transition}.
\newblock \bibinfo{journal}{Joule} \bibinfo{volume}{4},
  \bibinfo{pages}{2059--2064}.
\newblock \DOIprefix\doi{10.1016/j.joule.2020.07.022}.
\bibitem[{Schill et~al.(2015)Schill, Diekmann and Zerrahn}]{Schill_2015}
\bibinfo{author}{Schill, W.P.}, \bibinfo{author}{Diekmann, J.},
  \bibinfo{author}{Zerrahn, A.}, \bibinfo{year}{2015}.
\newblock \bibinfo{title}{Power storage: An important option for the german
  energy transition}.
\newblock \bibinfo{journal}{{DIW DIW Economic Bulletin}} \bibinfo{volume}{10},
  \bibinfo{pages}{137 -- 146}.
\bibitem[{Schill et~al.(2017a)Schill, Pahle and Gambardella}]{Schill_NE_2017}
\bibinfo{author}{Schill, W.P.}, \bibinfo{author}{Pahle, M.},
  \bibinfo{author}{Gambardella, C.}, \bibinfo{year}{2017}a.
\newblock \bibinfo{title}{Start-up costs of thermal power plants in markets
  with increasing shares of variable renewable generation}.
\newblock \bibinfo{journal}{Nature Energy} \bibinfo{volume}{2},
  \bibinfo{pages}{17050}.
\newblock \DOIprefix\doi{10.1038/nenergy.2017.50}.
\bibitem[{Schill and Zerrahn(2018)}]{Schill_2018}
\bibinfo{author}{Schill, W.P.}, \bibinfo{author}{Zerrahn, A.},
  \bibinfo{year}{2018}.
\newblock \bibinfo{title}{Long-run power storage requirements for high shares
  of renewables: Results and sensitivities}.
\newblock \bibinfo{journal}{Renewable and Sustainable Energy Reviews}
  \bibinfo{volume}{83}, \bibinfo{pages}{156--171}.
\newblock \DOIprefix\doi{10.1016/j.rser.2017.05.205}.
\bibitem[{Schill and Zerrahn(2020)}]{Schill_zenodo_2020}
\bibinfo{author}{Schill, W.P.}, \bibinfo{author}{Zerrahn, A.},
  \bibinfo{year}{2020}.
\newblock \bibinfo{title}{Reduced version of the model {DIETER}}.
\newblock \bibinfo{howpublished}{Zenodo}.
\newblock \DOIprefix\doi{10.5281/zenodo.4383288}.
\bibitem[{Schill et~al.(2017b)Schill, Zerrahn and Kunz}]{Schill_2017}
\bibinfo{author}{Schill, W.P.}, \bibinfo{author}{Zerrahn, A.},
  \bibinfo{author}{Kunz, F.}, \bibinfo{year}{2017}b.
\newblock \bibinfo{title}{Prosumage of solar electricity: Pros, cons, and the
  system perspective}.
\newblock \bibinfo{journal}{Economics of Energy \& Environmental Policy}
  \bibinfo{volume}{6}, \bibinfo{pages}{7 -- 31}.
\newblock \DOIprefix\doi{10.5547/2160-5890.6.1.wsch}.
\bibitem[{Schlachtberger et~al.(2017)Schlachtberger, Brown, Schramm and
  Greiner}]{Schlachtberger_2017}
\bibinfo{author}{Schlachtberger, D.}, \bibinfo{author}{Brown, T.},
  \bibinfo{author}{Schramm, S.}, \bibinfo{author}{Greiner, M.},
  \bibinfo{year}{2017}.
\newblock \bibinfo{title}{The benefits of cooperation in a highly renewable
  {European} electricity network}.
\newblock \bibinfo{journal}{Energy} \bibinfo{volume}{134}, \bibinfo{pages}{469
  -- 481}.
\newblock \DOIprefix\doi{10.1016/j.energy.2017.06.004}.
\bibitem[{Schmidt et~al.(2017)Schmidt, Hawkes, Gambhir and
  Staffell}]{Schmidt_2017}
\bibinfo{author}{Schmidt, O.}, \bibinfo{author}{Hawkes, A.},
  \bibinfo{author}{Gambhir, A.}, \bibinfo{author}{Staffell, I.},
  \bibinfo{year}{2017}.
\newblock \bibinfo{title}{The future cost of electrical energy storage based on
  experience rates}.
\newblock \bibinfo{journal}{Nature Energy} \bibinfo{volume}{2},
  \bibinfo{pages}{17110}.
\newblock \DOIprefix\doi{10.1038/nenergy.2017.110}.
\bibitem[{Schmidt et~al.(2019)Schmidt, Melchior, Hawkes and
  Staffell}]{Schmidt2019}
\bibinfo{author}{Schmidt, O.}, \bibinfo{author}{Melchior, S.},
  \bibinfo{author}{Hawkes, A.}, \bibinfo{author}{Staffell, I.},
  \bibinfo{year}{2019}.
\newblock \bibinfo{title}{{Projecting the Future Levelized Cost of Electricity
  Storage Technologies}}.
\newblock \bibinfo{journal}{Joule} \bibinfo{volume}{3},
  \bibinfo{pages}{81--100}.
\newblock \DOIprefix\doi{10.1016/j.joule.2018.12.008}.
\bibitem[{Seel et~al.(2018)Seel, Mills, Wiser, Deb, Asokkumar, Hassanzadeh and
  Aarabali}]{Seel2018}
\bibinfo{author}{Seel, J.}, \bibinfo{author}{Mills, A.},
  \bibinfo{author}{Wiser, R.}, \bibinfo{author}{Deb, S.},
  \bibinfo{author}{Asokkumar, A.}, \bibinfo{author}{Hassanzadeh, M.},
  \bibinfo{author}{Aarabali, A.}, \bibinfo{year}{2018}.
\newblock \bibinfo{title}{{Impacts of High Variable Renewable Energy Futures on
  Wholesale Electricity Prices, and on Electric-Sector Decision Making}}.
\newblock \bibinfo{type}{Technical Report} \bibinfo{number}{May}.
\newblock \URLprefix
  \url{https://emp.lbl.gov/publications/impacts-high-variable-renewable}.
\bibitem[{Sensfu{\ss} et~al.(2008)Sensfu{\ss}, Ragwitz and
  Genoese}]{Sensfuss2008}
\bibinfo{author}{Sensfu{\ss}, F.}, \bibinfo{author}{Ragwitz, M.},
  \bibinfo{author}{Genoese, M.}, \bibinfo{year}{2008}.
\newblock \bibinfo{title}{{The merit-order effect: A detailed analysis of the
  price effect of renewable electricity generation on spot market prices in
  Germany}}.
\newblock \bibinfo{journal}{Energy Policy} \bibinfo{volume}{36},
  \bibinfo{pages}{3076--3084}.
\newblock \DOIprefix\doi{10.1016/j.enpol.2008.03.035}.
\bibitem[{Sinn(2017)}]{Sinn_2017}
\bibinfo{author}{Sinn, H.W.}, \bibinfo{year}{2017}.
\newblock \bibinfo{title}{Buffering volatility: A study on the limits of
  {G}ermany's energy revolution}.
\newblock \bibinfo{journal}{European Economic Review} \bibinfo{volume}{99},
  \bibinfo{pages}{130 -- 150}.
\newblock \DOIprefix\doi{10.1016/j.euroecorev.2017.05.007}.
\bibitem[{Sioshansi et~al.(2012)Sioshansi, Denholm and Jenkin}]{Sioshansi_2012}
\bibinfo{author}{Sioshansi, R.}, \bibinfo{author}{Denholm, P.},
  \bibinfo{author}{Jenkin, T.}, \bibinfo{year}{2012}.
\newblock \bibinfo{title}{Market and policy barriers to deployment of energy
  storage}.
\newblock \bibinfo{journal}{Economics of Energy \& Environmental Policy}
  \bibinfo{volume}{1}, \bibinfo{pages}{47--64}.
\bibitem[{Sioshansi et~al.(2009)Sioshansi, Denholm, Jenkin and
  Weiss}]{Sioshansi_2009}
\bibinfo{author}{Sioshansi, R.}, \bibinfo{author}{Denholm, P.},
  \bibinfo{author}{Jenkin, T.}, \bibinfo{author}{Weiss, J.},
  \bibinfo{year}{2009}.
\newblock \bibinfo{title}{Estimating the value of electricity storage in {PJM}:
  Arbitrage and some welfare effects}.
\newblock \bibinfo{journal}{Energy Economics} \bibinfo{volume}{31},
  \bibinfo{pages}{269 -- 277}.
\newblock \DOIprefix\doi{10.1016/j.eneco.2008.10.005}.
\bibitem[{de~Sisternes et~al.(2016)de~Sisternes, Jenkins and
  Botterud}]{deSisternes_2016}
\bibinfo{author}{de~Sisternes, F.J.}, \bibinfo{author}{Jenkins, J.D.},
  \bibinfo{author}{Botterud, A.}, \bibinfo{year}{2016}.
\newblock \bibinfo{title}{The value of energy storage in decarbonizing the
  electricity sector}.
\newblock \bibinfo{journal}{Applied Energy} \bibinfo{volume}{175},
  \bibinfo{pages}{368 -- 379}.
\newblock \DOIprefix\doi{10.1016/j.apenergy.2016.05.014}.
\bibitem[{Staffell and Rustomji(2016)}]{Staffell_2016}
\bibinfo{author}{Staffell, I.}, \bibinfo{author}{Rustomji, M.},
  \bibinfo{year}{2016}.
\newblock \bibinfo{title}{Maximising the value of electricity storage}.
\newblock \bibinfo{journal}{Journal of Energy Storage} \bibinfo{volume}{8},
  \bibinfo{pages}{212 -- 225}.
\newblock \DOIprefix\doi{10.1016/j.est.2016.08.010}.
\bibitem[{Steinke et~al.(2013)Steinke, Wolfrum and Hoffmann}]{Steinke_2013}
\bibinfo{author}{Steinke, F.}, \bibinfo{author}{Wolfrum, P.},
  \bibinfo{author}{Hoffmann, C.}, \bibinfo{year}{2013}.
\newblock \bibinfo{title}{{Grid vs. storage in a 100\% renewable Europe}}.
\newblock \bibinfo{journal}{Renewable Energy} \bibinfo{volume}{50},
  \bibinfo{pages}{826 -- 832}.
\newblock \DOIprefix\doi{10.1016/j.renene.2012.07.044}.
\bibitem[{Stephan et~al.(2016)Stephan, Battke, Beuse, Clausdeinken and
  Schmidt}]{Stephan_2016}
\bibinfo{author}{Stephan, A.}, \bibinfo{author}{Battke, B.},
  \bibinfo{author}{Beuse, M.}, \bibinfo{author}{Clausdeinken, J.},
  \bibinfo{author}{Schmidt, T.}, \bibinfo{year}{2016}.
\newblock \bibinfo{title}{Limiting the public cost of stationary battery
  deployment by combining applications}.
\newblock \bibinfo{journal}{Nature Energy} \bibinfo{volume}{1},
  \bibinfo{pages}{16079}.
\newblock \DOIprefix\doi{10.1038/nenergy.2016.79}.
\bibitem[{Stoft(2002)}]{Stoft2002}
\bibinfo{author}{Stoft, S.}, \bibinfo{year}{2002}.
\newblock \bibinfo{title}{{Power System Economics: Designing Markets for
  Electricity}}.
\newblock \URLprefix
  \url{https://www.wiley.com/en-us/Power+System+Economics%3A+Designing+Markets+for+Electricity-p-9780471150404}.
\bibitem[{Taljegard et~al.(2019)Taljegard, Walter, G\"{o}ransson, Odenberger
  and Johnsson}]{Taljegard_2019}
\bibinfo{author}{Taljegard, M.}, \bibinfo{author}{Walter, V.},
  \bibinfo{author}{G\"{o}ransson, L.}, \bibinfo{author}{Odenberger, M.},
  \bibinfo{author}{Johnsson, F.}, \bibinfo{year}{2019}.
\newblock \bibinfo{title}{Impact of electric vehicles on the
  cost-competitiveness of generation and storage technologies in the
  electricity system}.
\newblock \bibinfo{journal}{Environmental Research Letters}
  \bibinfo{volume}{14}, \bibinfo{pages}{124087}.
\newblock \DOIprefix\doi{10.1088/1748-9326/ab5e6b}.
\bibitem[{Tol(2009)}]{Tol2009}
\bibinfo{author}{Tol, R.S.J.}, \bibinfo{year}{2009}.
\newblock \bibinfo{title}{{The Economie Effects of Climate Change}}.
\newblock \bibinfo{journal}{J. Econ. Perspect.} \bibinfo{volume}{23},
  \bibinfo{pages}{29--51}.
\newblock \DOIprefix\doi{10.1257/jep.23.2.29}.
\bibitem[{Tong et~al.(2020)Tong, Yuan, Lewis, Davis and Caldeira}]{Tong_2020}
\bibinfo{author}{Tong, F.}, \bibinfo{author}{Yuan, M.}, \bibinfo{author}{Lewis,
  N.S.}, \bibinfo{author}{Davis, S.J.}, \bibinfo{author}{Caldeira, K.},
  \bibinfo{year}{2020}.
\newblock \bibinfo{title}{Effects of deep reductions in energy storage costs on
  highly reliable wind and solar electricity systems}.
\newblock \bibinfo{journal}{iScience} \bibinfo{volume}{23},
  \bibinfo{pages}{101484}.
\newblock \DOIprefix\doi{10.1016/j.isci.2020.101484}.
\bibitem[{Ueckerdt et~al.(2015)Ueckerdt, Brecha, Luderer, Sullivan, Schmid,
  Bauer, B\"{o}ttger and Pietzcker}]{Ueckerdt_2015}
\bibinfo{author}{Ueckerdt, F.}, \bibinfo{author}{Brecha, R.},
  \bibinfo{author}{Luderer, G.}, \bibinfo{author}{Sullivan, P.},
  \bibinfo{author}{Schmid, E.}, \bibinfo{author}{Bauer, N.},
  \bibinfo{author}{B\"{o}ttger, D.}, \bibinfo{author}{Pietzcker, R.},
  \bibinfo{year}{2015}.
\newblock \bibinfo{title}{Representing power sector variability and the
  integration of variable renewables in long-term energy-economy models using
  residual load duration curves}.
\newblock \bibinfo{journal}{Energy} \bibinfo{volume}{90}, \bibinfo{pages}{1799
  -- 1814}.
\newblock \DOIprefix\doi{10.1016/j.energy.2015.07.006}.
\bibitem[{Ueckerdt et~al.(2013)Ueckerdt, Hirth, Luderer and
  Edenhofer}]{Ueckerdt2013a}
\bibinfo{author}{Ueckerdt, F.}, \bibinfo{author}{Hirth, L.},
  \bibinfo{author}{Luderer, G.}, \bibinfo{author}{Edenhofer, O.},
  \bibinfo{year}{2013}.
\newblock \bibinfo{title}{{System LCOE: What are the costs of variable
  renewables?}}
\newblock \bibinfo{journal}{Energy} \bibinfo{volume}{63},
  \bibinfo{pages}{61--75}.
\newblock \DOIprefix\doi{10.1016/j.energy.2013.10.072}.
\bibitem[{Vartiainen et~al.(2020)Vartiainen, Masson, Breyer, Moser and
  {Rom{\'{a}}n Medina}}]{Vartiainen2020}
\bibinfo{author}{Vartiainen, E.}, \bibinfo{author}{Masson, G.},
  \bibinfo{author}{Breyer, C.}, \bibinfo{author}{Moser, D.},
  \bibinfo{author}{{Rom{\'{a}}n Medina}, E.}, \bibinfo{year}{2020}.
\newblock \bibinfo{title}{{Impact of weighted average cost of capital, capital
  expenditure, and other parameters on future utility-scale PV levelised cost
  of electricity}}.
\newblock \bibinfo{journal}{Prog. Photovoltaics Res. Appl.}
  \bibinfo{volume}{28}, \bibinfo{pages}{439--453}.
\newblock \DOIprefix\doi{10.1002/pip.3189}.
\bibitem[{Weitemeyer et~al.(2015)Weitemeyer, Kleinhans, Vogt and
  Agert}]{Weitemeyer_2015}
\bibinfo{author}{Weitemeyer, S.}, \bibinfo{author}{Kleinhans, D.},
  \bibinfo{author}{Vogt, T.}, \bibinfo{author}{Agert, C.},
  \bibinfo{year}{2015}.
\newblock \bibinfo{title}{Integration of renewable energy sources in future
  power systems: The role of storage}.
\newblock \bibinfo{journal}{Renewable Energy} \bibinfo{volume}{75},
  \bibinfo{pages}{14 -- 20}.
\newblock \DOIprefix\doi{10.1016/j.renene.2014.09.028}.
\bibitem[{Welisch et~al.(2016)Welisch, Ortner and Resch}]{Welisch2016}
\bibinfo{author}{Welisch, M.}, \bibinfo{author}{Ortner, A.},
  \bibinfo{author}{Resch, G.}, \bibinfo{year}{2016}.
\newblock \bibinfo{title}{{Assessment of RES technology market values and the
  merit-order effect - An econometric multi-country analysis}}.
\newblock \bibinfo{journal}{Energy Environ.} \bibinfo{volume}{27},
  \bibinfo{pages}{105--121}.
\newblock \DOIprefix\doi{10.1177/0958305X16638574}.
\bibitem[{Wenders(1976)}]{Wenders1976}
\bibinfo{author}{Wenders, J.T.}, \bibinfo{year}{1976}.
\newblock \bibinfo{title}{{Peak load pricing in the electric utility
  industry}}.
\newblock \bibinfo{journal}{Bell J. Econ.} \bibinfo{volume}{7},
  \bibinfo{pages}{232--241}.
\newblock \DOIprefix\doi{10.2307/3003198}.
\bibitem[{Wiese et~al.(2019)Wiese, Schlecht, Bunke, Gerbaulet, Hirth, Jahn,
  Kunz, Lorenz, M{\"{u}}hlenpfordt, Reimann and Schill}]{Wiese2019}
\bibinfo{author}{Wiese, F.}, \bibinfo{author}{Schlecht, I.},
  \bibinfo{author}{Bunke, W.D.}, \bibinfo{author}{Gerbaulet, C.},
  \bibinfo{author}{Hirth, L.}, \bibinfo{author}{Jahn, M.},
  \bibinfo{author}{Kunz, F.}, \bibinfo{author}{Lorenz, C.},
  \bibinfo{author}{M{\"{u}}hlenpfordt, J.}, \bibinfo{author}{Reimann, J.},
  \bibinfo{author}{Schill, W.P.}, \bibinfo{year}{2019}.
\newblock \bibinfo{title}{{Open Power System Data -- Frictionless data for
  electricity system modelling}}.
\newblock \bibinfo{journal}{Appl. Energy} \bibinfo{volume}{236},
  \bibinfo{pages}{401--409}.
\newblock \DOIprefix\doi{10.1016/j.apenergy.2018.11.097}.
\bibitem[{Woo et~al.(2011)Woo, Horowitz, Moore and Pacheco}]{Woo2011}
\bibinfo{author}{Woo, C.K.}, \bibinfo{author}{Horowitz, I.},
  \bibinfo{author}{Moore, J.}, \bibinfo{author}{Pacheco, A.},
  \bibinfo{year}{2011}.
\newblock \bibinfo{title}{{The impact of wind generation on the electricity
  spot-market price level and variance: The Texas experience}}.
\newblock \bibinfo{journal}{Energy Policy} \bibinfo{volume}{39},
  \bibinfo{pages}{3939--3944}.
\newblock \DOIprefix\doi{10.1016/j.enpol.2011.03.084}.
\bibitem[{Woo et~al.(2016)Woo, Moore, Schneiderman, Ho, Olson, Alagappan,
  Chawla, Toyama and Zarnikau}]{Woo2016}
\bibinfo{author}{Woo, C.K.}, \bibinfo{author}{Moore, J.},
  \bibinfo{author}{Schneiderman, B.}, \bibinfo{author}{Ho, T.},
  \bibinfo{author}{Olson, A.}, \bibinfo{author}{Alagappan, L.},
  \bibinfo{author}{Chawla, K.}, \bibinfo{author}{Toyama, N.},
  \bibinfo{author}{Zarnikau, J.}, \bibinfo{year}{2016}.
\newblock \bibinfo{title}{{Merit-order effects of renewable energy and price
  divergence in California's day-ahead and real-time electricity markets}}.
\newblock \bibinfo{journal}{Energy Policy} \bibinfo{volume}{92},
  \bibinfo{pages}{299--312}.
\newblock \DOIprefix\doi{10.1016/j.enpol.2016.02.023}.
\bibitem[{{World Energy Council}(2016)}]{WEC2016}
\bibinfo{author}{{World Energy Council}}, \bibinfo{year}{2016}.
\newblock \bibinfo{title}{{E-storage : Shifting from cost to value. Wind and
  solar applications}}.
\newblock \bibinfo{type}{World Energy Resources}.
\newblock \URLprefix
  \url{https://www.worldenergy.org/assets/downloads/Resources-E-storage-report-2016.02.04.pdf}.
\bibitem[{Yan et~al.(2019)Yan, Hitt, Turner and Mallouk}]{Yan_2019}
\bibinfo{author}{Yan, Z.}, \bibinfo{author}{Hitt, J.L.},
  \bibinfo{author}{Turner, J.A.}, \bibinfo{author}{Mallouk, T.E.},
  \bibinfo{year}{2019}.
\newblock \bibinfo{title}{{R}enewable {E}lectricity {S}torage {U}sing
  {E}lectrolysis}.
\newblock \bibinfo{journal}{Proceedings of the National Academy of Sciences}
  \bibinfo{volume}{117}, \bibinfo{pages}{12558--12563}.
\newblock \DOIprefix\doi{10.1073/pnas.1821686116}.
\bibitem[{Zerrahn and Schill(2017)}]{Zerrahn_2017}
\bibinfo{author}{Zerrahn, A.}, \bibinfo{author}{Schill, W.P.},
  \bibinfo{year}{2017}.
\newblock \bibinfo{title}{Long-run power storage requirements for high shares
  of renewables: review and a new model}.
\newblock \bibinfo{journal}{Renewable and Sustainable Energy Reviews}
  \bibinfo{volume}{79}, \bibinfo{pages}{1518 -- 1534}.
\newblock \DOIprefix\doi{10.1016/j.rser.2016.11.098}.
\bibitem[{Zerrahn et~al.(2018)Zerrahn, Schill and Kemfert}]{Zerrahn_2018}
\bibinfo{author}{Zerrahn, A.}, \bibinfo{author}{Schill, W.P.},
  \bibinfo{author}{Kemfert, C.}, \bibinfo{year}{2018}.
\newblock \bibinfo{title}{On the economics of electrical storage for variable
  renewable energy sources}.
\newblock \bibinfo{journal}{European Economic Review} \bibinfo{volume}{108},
  \bibinfo{pages}{259 -- 279}.
\newblock \DOIprefix\doi{10.1016/j.euroecorev.2018.07.004}.
\bibitem[{Ziegler et~al.(2019)Ziegler, Mueller, Pereira, Song, Ferrara, Chiang
  and Trancik}]{Ziegler_2019}
\bibinfo{author}{Ziegler, M.S.}, \bibinfo{author}{Mueller, J.M.},
  \bibinfo{author}{Pereira, G.D.}, \bibinfo{author}{Song, J.},
  \bibinfo{author}{Ferrara, M.}, \bibinfo{author}{Chiang, Y.M.},
  \bibinfo{author}{Trancik, J.E.}, \bibinfo{year}{2019}.
\newblock \bibinfo{title}{Storage requirements and costs of shaping renewable
  energy toward grid decarbonization}.
\newblock \bibinfo{journal}{Joule} \bibinfo{volume}{3}, \bibinfo{pages}{2134 --
  2153}.
\newblock \DOIprefix\doi{10.1016/j.joule.2019.06.012}.
\bibitem[{Zipp(2017)}]{Zipp2017}
\bibinfo{author}{Zipp, A.}, \bibinfo{year}{2017}.
\newblock \bibinfo{title}{{The marketability of variable renewable energy in
  liberalized electricity markets – An empirical analysis}}.
\newblock \bibinfo{journal}{Renew. Energy} \bibinfo{volume}{113},
  \bibinfo{pages}{1111--1121}.
\newblock \DOIprefix\doi{10.1016/j.renene.2017.06.072}.

\end{thebibliography}

\end{document}